\documentclass[%
reprint,
superscriptaddress,
showpacs,
 amsmath,amssymb,
 aps,prl,
floatfix,
]{revtex4-2}
\usepackage[english]{babel}
\usepackage[utf8]{inputenc}
\usepackage{xcolor}
\usepackage{amsthm}
\usepackage{textcomp}
\usepackage{mathtools}
\usepackage{multirow}
\usepackage{graphicx}
\usepackage{adjustbox}
\usepackage{placeins}
\usepackage[T1]{fontenc}
\usepackage{lipsum}
\usepackage{csquotes}
\usepackage[pdftex, pdftitle={Article}, pdfauthor={Author},colorlinks = true,allcolors = blue]{hyperref} 
\usepackage{dcolumn}
\usepackage{bm}
\usepackage{comment}
\usepackage{blindtext}
\usepackage{hyperref}
\hypersetup{
     colorlinks   = true,
     citecolor    = blue,
     linkcolor    = blue,
     urlcolor     = blue
}
\usepackage{etoolbox}
\usepackage[mathlines]{lineno}

\begin{document}
\title{Suppression of neutral pion production in deep-inelastic scattering off nuclei with the CLAS detector}

\newcommand*{\ULS}{Universidad de La Serena, Avda. Juan Cisternas 1200, La Serena, Chile}
\newcommand*{\ULSindex}{1}
\affiliation{\ULS}
\newcommand*{\CCTVal}{Center for Science and Technology of Valpara\'iso 699, Valpara\'iso, Chile}
\newcommand*{\CCTValindex}{3}
\affiliation{\CCTVal}
\newcommand*{\UTFSM}{Universidad T\'{e}cnica Federico Santa Mar\'{i}a, Casilla, 110-V Valpara\'{i}so, Chile}
\newcommand*{\UTFSMindex}{2}
\affiliation{\UTFSM}
\newcommand*{\SAPHIR}{SAPHIR Millennium Science Institute, Santiago, Chile}
\newcommand*{\SAPHIRindex}{4}
\affiliation{\SAPHIR}
\newcommand*{\JLAB}{Thomas Jefferson National Accelerator Facility, Newport News, VA 23606}
\newcommand*{\JLABindex}{5}
\affiliation{\JLAB}
\newcommand*{\DUKE}{Duke University, Durham, NC 27708}
\newcommand*{\DUKEindex}{5}
\affiliation{\DUKE}
\newcommand*{\UCONN}{University of Connecticut, Storrs, CT 06269}
\newcommand*{\UCONNindex}{6}
\affiliation{\UCONN}
\newcommand*{\Serena}{Universidad de La Serena,  1720170 La Serena, Chile}
\newcommand*{\Serenaindex}{7}
\affiliation{\Serena}
\newcommand*{\ANL}{Argonne National Laboratory, Argonne, IL 60439}
\newcommand*{\ANLindex}{8}
\affiliation{\ANL}
\newcommand*{\CSUDH}{California State University, Dominguez Hills, Carson, CA 90747}
\newcommand*{\CSUDHindex}{9}
\affiliation{\CSUDH}
\newcommand*{\CANISIUS}{Canisius College, Buffalo, NY 14208}
\newcommand*{\CANISIUSindex}{10}
\affiliation{\CANISIUS}
\newcommand*{\CMU}{Carnegie Mellon University, Pittsburgh, PA 15213}
\newcommand*{\CMUindex}{11}
\affiliation{\CMU}
\newcommand*{\CUA}{Catholic University of America, Washington, D.C. 20064}
\newcommand*{\CUAindex}{12}
\affiliation{\CUA}
\newcommand*{\SACLAY}{IRFU, CEA, Universit\'{e} Paris-Saclay, F-91191 Gif-sur-Yvette, France}
\newcommand*{\SACLAYindex}{13}
\affiliation{\SACLAY}
\newcommand*{\CNU}{Christopher Newport University, Newport News, VA 23606}
\newcommand*{\CNUindex}{14}
\affiliation{\CNU}
\newcommand*{\DUQUESNE}{Duquesne University, 600 Forbes Avenue, Pittsburgh, PA 15282 }
\newcommand*{\DUQUESNEindex}{15}
\affiliation{\DUQUESNE}
\newcommand*{\FU}{Fairfield University, Fairfield, CT 06824}
\newcommand*{\FUindex}{16}
\affiliation{\FU}
\newcommand*{\FERRARAU}{Universit\`{a} di Ferrara, 44121 Ferrara, Italy}
\newcommand*{\FERRARAUindex}{17}
\affiliation{\FERRARAU}
\newcommand*{\FIU}{Florida International University, Miami, FL 33199}
\newcommand*{\FIUindex}{18}
\affiliation{\FIU}
\newcommand*{\FSU}{Florida State University, Tallahassee, FL 32306}
\newcommand*{\FSUindex}{19}
\affiliation{\FSU}
\newcommand*{\GWUI}{The George Washington University, Washington, D.C. 20052}
\newcommand*{\GWUIindex}{20}
\affiliation{\GWUI}
\newcommand*{\GSIFFN}{GSI Helmholtzzentrum f\"ur Schwerionenforschung GmbH, D-64291
Darmstadt, Germany}
\newcommand*{\GSIFFNindex}{21}
\affiliation{\GSIFFN}
\newcommand*{\INFNFE}{INFN, Sezione di Ferrara, 44100 Ferrara, Italy}
\newcommand*{\INFNFEindex}{22}
\affiliation{\INFNFE}
\newcommand*{\INFNFR}{INFN, Laboratori Nazionali di Frascati, 00044 Frascati, Italy}
\newcommand*{\INFNFRindex}{23}
\affiliation{\INFNFR}
\newcommand*{\INFNGE}{INFN, Sezione di Genova, 16146 Genova, Italy}
\newcommand*{\INFNGEindex}{24}
\affiliation{\INFNGE}
\newcommand*{\INFNRO}{INFN, Sezione di Roma Tor Vergata, 00133 Rome, Italy}
\newcommand*{\INFNROindex}{25}
\affiliation{\INFNRO}
\newcommand*{\INFNTUR}{INFN, Sezione di Torino, 10125 Torino, Italy}
\newcommand*{\INFNTURindex}{26}
\affiliation{\INFNTUR}
\newcommand*{\INFNPAV}{INFN, Sezione di Pavia, 27100 Pavia, Italy}
\newcommand*{\INFNPAVindex}{27}
\affiliation{\INFNPAV}
\newcommand*{\ORSAY}{Universit'{e} Paris-Saclay, CNRS/IN2P3, IJCLab, 91405 Orsay, France}
\newcommand*{\ORSAYindex}{28}
\affiliation{\ORSAY}
\newcommand*{\Juelich}{Institut f\"ur Kernphysik (Juelich), Juelich, 52428, Germany}
\newcommand*{\Juelichindex}{29}
\affiliation{\Juelich}
\newcommand*{\JMU}{James Madison University, Harrisonburg, VA 22807}
\newcommand*{\JMUindex}{30}
\affiliation{\JMU}
\newcommand*{\KNU}{Kyungpook National University, Daegu 41566, Republic of Korea}
\newcommand*{\KNUindex}{31}
\affiliation{\KNU}
\newcommand*{\LAMAR}{Lamar University, 4400 MLK Blvd, PO Box 10046, Beaumont, TX 77710}
\newcommand*{\LAMARindex}{32}
\affiliation{\LAMAR}
\newcommand*{\MIT}{Massachusetts Institute of Technology, Cambridge, MA  02139-4307}
\newcommand*{\MITindex}{33}
\affiliation{\MIT}
\newcommand*{\MISS}{Mississippi State University, Mississippi State, MS 39762-5167}
\newcommand*{\MISSindex}{34}
\affiliation{\MISS}
\newcommand*{\BRIN}{National Research and Innovation Agency (BRIN), Indonesia}
\newcommand*{\BRINindex}{35}
\affiliation{\BRIN}
\newcommand*{\ITEP}{National Research Centre Kurchatov Institute - ITEP, Moscow, 117259, Russia}
\newcommand*{\ITEPindex}{36}
\affiliation{\ITEP}
\newcommand*{\UNH}{University of New Hampshire, Durham, NH03824-3568}
\newcommand*{\UNHindex}{37}
\affiliation{\UNH}
\newcommand*{\NMSU}{New Mexico State University, PO Box 30001, Las Cruces, NM 88003}
\newcommand*{\NMSUindex}{38}
\affiliation{\NMSU}
\newcommand*{\NSU}{Norfolk State University, Norfolk, VA 23504}
\newcommand*{\NSUindex}{39}
\affiliation{\NSU}
\newcommand*{\OHIOU}{Ohio University, Athens, OH  45701}
\newcommand*{\OHIOUindex}{40}
\affiliation{\OHIOU}
\newcommand*{\ODU}{Old Dominion University, Norfolk, VA 23529}
\newcommand*{\ODUindex}{41}
\affiliation{\ODU}
\newcommand*{\JLUGiessen}{II Physikalisches Institut der Universitaet Giessen, 35392 Giessen, Germany}
\newcommand*{\JLUGiessenindex}{42}
\affiliation{\JLUGiessen}
\newcommand*{\RPI}{Rensselaer Polytechnic Institute, Troy, NY 12180-3590}
\newcommand*{\RPIindex}{43}
\affiliation{\RPI}
\newcommand*{\URICH}{University of Richmond, Richmond, VA 23173}
\newcommand*{\URICHindex}{44}
\affiliation{\URICH}
\newcommand*{\ROMAII}{Universit\`{a} di Roma Tor Vergata, 00133 Rome Italy}
\newcommand*{\ROMAIIindex}{45}
\affiliation{\ROMAII}
\newcommand*{\MSU}{Skobeltsyn Institute of Nuclear Physics, Lomonosov Moscow State University, 119234 Moscow, Russia}
\newcommand*{\MSUindex}{46}
\affiliation{\MSU}
\newcommand*{\SCAROLINA}{University of South Carolina, Columbia, SC 29208}
\newcommand*{\SCAROLINAindex}{47}
\affiliation{\SCAROLINA}
\newcommand*{\TEMPLE}{Temple University,  Philadelphia, PA 19122 }
\newcommand*{\TEMPLEindex}{48}
\affiliation{\TEMPLE}
\newcommand*{\INSUBRIA}{Universit\`{a} degli Studi dell'Insubria, 22100 Como, Italy}
\newcommand*{\INSUBRIAindex}{49}
\affiliation{\INSUBRIA}
\newcommand*{\BRESCIA}{Universit\`{a} degli Studi di Brescia, 25123 Brescia, Italy}
\newcommand*{\BRESCIAindex}{50}
\affiliation{\BRESCIA}
\newcommand*{\UCR}{University of California Riverside, 900 University Avenue, Riverside, CA 92521}
\newcommand*{\UCRindex}{51}
\affiliation{\UCR}
\newcommand*{\GLASGOW}{University of Glasgow, Glasgow G12 8QQ, United Kingdom}
\newcommand*{\GLASGOWindex}{52}
\affiliation{\GLASGOW}
\newcommand*{\YORK}{University of York, York YO10 5DD, United Kingdom}
\newcommand*{\YORKindex}{53}
\affiliation{\YORK}
\newcommand*{\VIRGINIA}{University of Virginia, Charlottesville, VA 22901}
\newcommand*{\VIRGINIAindex}{54}
\affiliation{\VIRGINIA}
\newcommand*{\WM}{College of William and Mary, Williamsburg, VA 23187-8795}
\newcommand*{\WMindex}{55}
\affiliation{\WM}
\newcommand*{\YEREVAN}{Yerevan Physics Institute, 375036 Yerevan, Armenia}
\newcommand*{\YEREVANindex}{56}
\affiliation{\YEREVAN}
\newcommand*{\NOWISU}{Idaho State University, Pocatello, Idaho 83209}
\newcommand*{\NOWISUindex}{57}
\affiliation{\NOWISU}
\author {T.~Mineeva} 
\email{mineeva@jlab.org}
\affiliation{\ULS}
\affiliation{\CCTVal}
\author {W.~K.~Brooks} 
\affiliation{\UTFSM}
\affiliation{\CCTVal}
\affiliation{\SAPHIR}
\affiliation{\JLAB}
\affiliation{\DUKE}

\author{A. El Alaoui}
\affiliation{\UTFSM}
\affiliation{\CCTVal}
\author{H. Hakobyan}
\affiliation{\UTFSM}
\affiliation{\CCTVal}
\author{K. Joo}
\affiliation{\UCONN}
\author{J.A. L\'opez}
\affiliation{\UTFSM}
\author{O. Soto}
\affiliation{\Serena}

\author {P.~Achenbach} 
\affiliation{\JLAB}
\author {Z.~Akbar} 
\affiliation{\BRIN}
\author {J. S. Alvarado} 
\affiliation{\ORSAY}
% ...
\collaboration{The CLAS Collaboration}
\noaffiliation
\date{\today}

\begin{abstract}
We present the first three-fold differential measurement for neutral pion multiplicity ratios produced in semi-inclusive deep-inelastic electron scattering on carbon, iron and lead nuclei normalized to deuterium from CLAS  at Jefferson Lab. 
We found that the neutral pion multiplicity ratio is maximally suppressed for the leading hadrons (energy fraction $z\rightarrow$~1), suppression varying from 25\% in carbon up to 75\% in lead.
An enhancement of the multiplicity ratio at low $z$ and high $p_{T}^{2}$ is observed, suggesting an interconnection between these two variables.
This behavior is qualitatively similar to the  previous two-fold differential measurement of charged pions by the HERMES Collaboration and recently - by CLAS Collaboration.
The largest enhancement was observed at high $p_{T}^{2}$
for heavier nuclei, namely iron and lead, while the smallest enhancement was observed for the lightest nucleus, carbon. 
This behavior suggests a competition between partonic multiple scattering, which causes enhancement, and hadronic inelastic scattering, which causes suppression.
\end{abstract}
\maketitle

\section{I. INTRODUCTION}

Hadron formation is one of the last frontiers of QCD. While successful models of this 
process exist, they only have a tenuous connection to the underlying QCD origin of the 
process. 
The long distance scales involved in hadron formation currently preclude use of 
perturbative methods to calculate, for example, fragmentation functions (FF), which 
describe how color-carrying quarks and gluons turn into color-neutral hadrons or 
photons~\cite{METZ2016136}. \\
The kinematic region of lepton deep-inelastic scattering at high $x_{Bj}$, where 
$x_{Bj}$ is the fraction of the proton momentum carried by the struck quark, offers a 
powerfully simple interpretation compared to low $x_{Bj}$ where quark pair production 
dominates~\cite{PhysRevD.46.931}. 
In the single-photon exchange approximation, a valence quark absorbs the full energy 
and momentum of the virtual photon $\gamma^*$; thus, the energy transfer gives the 
initial energy of the struck 
quark, neglecting intrinsic quark momentum, and neglecting Fermi momentum of the 
nucleon for nuclear interactions.
At the same level of approximation, the initial direction of the struck quark is known from the momentum transfer of the collision,
which provides a unique reference axis.
For nuclear targets, this essentially creates a secondary  ``beam’’ of quarks of known energy and direction, for which the interaction with the nuclear system provides information at the femtometer distance scale.

An important experimental observable sensitive to the in-medium hadronization process - the complex process of the evolution of a struck quark into multiple hadrons - is the hadronic multiplicity ratio. It is defined as the normalized yield of hadron $h$ produced on a heavy nuclear target $A$ relative to a light nuclei, e.g., deuterium $D$:

\begin{equation}
R_{h}(\nu, Q^{2},z,p_{T}^{2})=\frac{N_{h}^{A}(\nu, Q^{2},z,p_{T}^{2})/N_{e}^{A}(\nu, Q^{2})}{N_{h}^{D}(\nu, Q^{2},z,p_{T}^{2})/N_{e}^{D}(\nu, Q^{2})},
\label{eq:R}
\end{equation}

\noindent where  $N_{h}$ is the number of hadrons produced in semi-inclusive deep-inelastic scattering (SIDIS) events, in which, following the $\gamma^*$ scattering off the quark, the leading hadron is detected in addition to the scattered electron;
 $N_e$ is the number of DIS electrons within the same inclusive kinematic bins for the 
 numerator as for the denominator; $Q^2$ is the $\gamma^*$ four-momentum transfer 
 squared, $\nu$ is the energy transferred which in the lab frame is defined as $\nu=E-
 E'$ ($E$ and $E'$ is energy of the incoming and outgoing electrons, respectively), $z$ 
 is the energy fraction of the hadron defined as $z$~=~$E_{h}$/$\nu$, and $p_T^{2}$ is 
 the component of the hadron momentum squared transverse to the $\gamma^*$ direction; 
 the dependence on $\phi_{pq}$, the azimuthal angle of the hadron with respect to 
 the lepton plane, was integrated over. 
The hadronic multiplicity ratio quantifies the extent to which hadron production is 
enhanced or attenuated in nuclei compared to deuterium; in the absence of any nuclear 
effects, this observable is equal to unity.

Nuclear SIDIS experiments have been performed in fixed-target conditions in 
facilities such as SLAC, CERN (SPS), DESY (HERMES) and Jefferson Lab (CLAS).
The study of nuclear SIDIS with fully identified final state hadrons began with the 
HERMES program, which published a series of papers between 2001 and 
2011~\cite{5ea5598ca62d4b039b6f3ff52e9f0923,AIRAPETIAN200337,HERMES:2007,HERMES:2005mar,
HERMES:2009uge,HERMES:2011qjb}, opening an era of quantitative studies of color 
propagation and hadron formation using nuclei as spatial analyzers.
The one- and two-fold differential hadron production data off nuclei can be described 
with some level of success by 
models~\cite{Kopeliovich_2004,Guiot_2020,Gallmeister_2008,Falter_2006,Falter_2004,Falter:2003nx,Falter_2004a,Wang_2002,Osborne_2002,Chang_2014,PhysRevLett.99.152301,Zhang_2007,Wang_2002a,Kang_2016,Kopeliovich_2008,Brooks_2021} 
using two in-medium ingredients: (1) quark energy loss and (2) interactions of forming hadrons with the nuclear medium.
The final HERMES paper ~\cite{HERMES:2011qjb} underlines the importance of 
multi-differential cross sections, since charged-hadron multiplicity data displays nontrivial features that cannot be captured by a one-dimensional description, particularly for the baryons.
A comprehensive review can be found in Ref.~\cite{cite-key}.
One-, two- and three-fold differential measurements of $R_{h}$ for identified hadrons   
were reported by CLAS experiments ~\cite{Daniel:2011nq, PhysRevLett.130.142301,CLAS:2021jhm}.

This paper presents the first three-fold differential measurement of neutral pion multiplicity ratios in SIDIS kinematics.
Neutral pions are substantially more difficult to measure than charged pions mainly because of limited statistics and the presence of combinatorial backgrounds. 
Nevertheless, they are an essential measurement for completing the understanding of the isospin triplet state of pion. 
While having a much more limited range in $Q^2$ and $\nu$,
the integrated luminosity in the new data set is two orders of magnitude greater than that of HERMES, dramatically increasing the statistical accuracy of the measurement.
This allowed us to extend one-dimensional HERMES $\pi^0$ data measured up to mass number 131~\cite{HERMES:2007}, 
to three-dimensional data with mass numbers up to 208.

\section{II. EXPERIMENTAL SETUP AND DATA ANALYSIS}
\begin{figure*}[h!tb]
\centering
\includegraphics[width=0.95\textwidth]{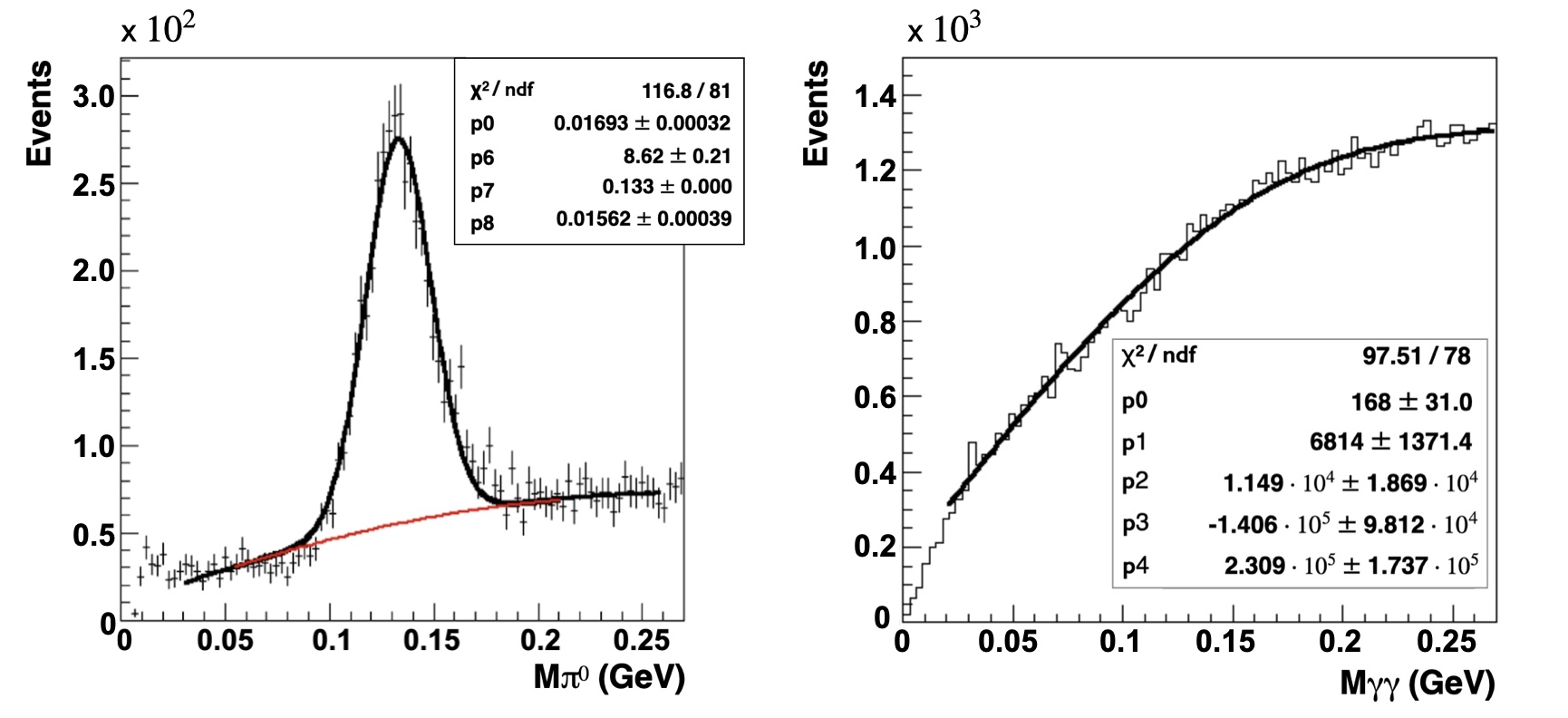}
\caption{
\small Left: Number of events as a function of $\pi^{0}$ invariant mass in a particular ($\nu$,~$z$,~$p_{T}^{2}$) bin with  fit to a scaled mixed background (red). Right: Number of events as a function of invariant mass of the corresponding mixed background fitted with a 4th order polynomial.  
The total fit function is:  p0$\cdot$(p1+p2$\cdot$x+p3$\cdot$x$^{2}$+p4$\cdot$x$^{3}$+p5$\cdot$x$^{4}$)+N$\cdot$exp$({\frac{-(x-\mu)^{2}}{2\cdot\sigma^{2}}})$, where  p$_0$ is background normalization, p$_1$-p$_5$ are predetermined by the event mixing, N is free parameter corresponding to normalization p6, $\mu$ and $\sigma$ are parameters p7 and p8, the latter three are reflected on the right plot. 
The fitting procedure was performed twice: 
first, in the range $0.03 < M_{\gamma\gamma} < 0.25$~GeV to provide an estimate to $\mu$ and $\sigma$;
second, in the range (-5$\sigma$,+5$\sigma$) as indicated by the length of the red curve. 
}
\label{fig:MassFit}
\end{figure*} 
The data were collected during the EG2 run period in Hall B of Jefferson Lab using the CEBAF Large Acceptance Spectrometer (CLAS) ~\cite{Mecking:2003zu} and a 5.014 GeV electron beam.
CLAS was based on a six-fold symmetric toroidal magnet, created by six large superconducting coils that divided the spectrometer into six independently instrumented sectors, and comprised of 
four types of detectors: drift chambers (DC) followed by Cerenkov counters (CC), time-of-flight (TOF) scintillators, and electromagnetic shower calorimeters (EC). Photons from $\pi^{0}$ decay were measured in the EC at angles from about 8 to 45 degrees. 

One key ingredient in reducing systematic uncertainties of the multiplicity ratios was the use of a dual-target. 
The target system consisting of a 2-cm-long liquid-deuterium cryotarget separated by 4~cm from independently insertable solid targets (see Ref.~\cite{HAKOBYAN2008218}).
The center of the cryotarget cell and the solid target were placed 30 cm and 25 cm upstream of the CLAS center, respectively, in order to increase acceptance for negatively charged particles.
The advantage of the double target is that since the electron beam passed simultaneously through both targets, time-dependent systematic effects were reduced.
A wealth of information was collected during EG2 experiment providing data for hadronization, color transparency~\cite{ElFassi:2012} and short-range correlations~\cite{Hen_2013} studies.

The SIDIS reaction $e+A \rightarrow e'+\pi^{0}+X$ is measured, where
$e$ and $e'$ are the incident and scattered electrons, respectively,
and $X$ is the undetected part of the hadronic final state.
Since the $\pi^{0}$ decays almost instantaneously into two photons ($\pi^{0}\rightarrow\gamma\gamma$), events with one scattered electron and at least two photons were selected.
The invariant mass of the two-photon system was used to identify $\pi^{0}$ candidates.

The scattered electrons were selected in the following ranges: 1.0~$<Q^2<$~4.1~GeV$^{2}$, 2.2~$<\nu<$~4.25~GeV and $W>$~2~GeV, where $W$ is $\gamma^*$-nucleon invariant mass squared. 
The requirement of $Q^2>$~1~GeV$^{2}$ and $W>$~2~GeV allowed to probe nucleon structure in the DIS regime and reduce nucleon resonance region contributions; 
the requirement of $\nu<$~4.25~GeV allowed to reduce the size of radiative effects reflected in the $y=\frac{\nu}{E}<$0.85 cut, where $y$ is the energy fraction of the $\gamma^*$.
These cuts ensured $x_{Bj}>$ 0.1, meaning that valence quarks in the target nucleon were probed.
Detector acceptance and experimental statistics dictated $\pi^{0}$ kinematics of 0.3~$<z<$~1.0 and 0~$<p_{T}^{2}<$~1.5~GeV$^{2}$.
The event phase space was divided into two sets of three-fold differential multiplicity ratios with: 1) a total of 108 bins in  ($\nu$,~$z$,~$p_{T}^{2}$) integrated over $Q^{2}$, and 2) a total of 54 bins  in ($Q^{2}$,~$\nu$,~$z$) integrated over $p_{T}^{2}$. 

The electron selection was done as following: first, a negatively charged track in the DC plus a signal in the TOF and EC was required;
next, this candidate must have matching between mirror number and projectile angle of the track in CC (this requirement is similar to the cut on the number of photoelectrons without removing good electrons);
it further must satisfy sampling fraction cut and have a minimum 
 energy deposited in EC and, lastly, satisfy a coincidence time cut between the EC and TOF signals.
 We excluded DC regions with non-uniform tracking efficiency and transverse shower leakage. 
In order to determine the origin of the scattering event, the intersection of the electron track with the plane containing the ideal beam position was used. However, during the experiment, the beam was offset introducing sector-dependent effects in the vertex reconstruction.
Electron-proton elastic scattering was used to determine the beam offset which was then used to correct the reconstructed interaction vertex for each event to make it sector-independent. 

Following electron identification, all the neutral hits were  considered in the EC
provided their energy exceeded 0.3 GeV.
Photons were separated from neutrons based on expected photon arrival time $\Delta t ~= ~t_{EC}$ ~-~$ l_{EC}/c$~-~$t_{start}$, where $t_{EC}$ is time at the EC, $l_{EC}$ is the distance from the target to the EC hit, $c$ is the speed of light and $t_{start}$ is the event start time ~\cite{PhD:Mineeva}.
To avoid transverse shower energy leakage,  events at the edge of the EC were rejected.  
 Photons detected within 12$^{\circ}$ of the electron track were  rejected in order to remove events from bremsstrahlung radiation. 
In order to improve $\pi^{0}$ resolution, measured photon energy was corrected for a small momentum dependence of the EC sampling fraction~\cite{PhD:Mineeva}.
Finally, $\pi^{0}$ candidates were reconstructed from all pairs of photons detected in each event (see Fig.~\ref{fig:MassFit}).  
After photon energy correction, the  minimum energy of $\pi^{0}$ candidate was $E_{\pi^{0}}>$0.5~GeV. 

To calculate the number of $\pi^{0}$'s, the two-photon invariant mass spectrum was fit with a Gaussian  function plus a polynomial background (see Fig.~\ref{fig:MassFit}). 
To determine the shape of combinatorial background,
an event mixing technique, consisted of combining photons from uncorrelated events was used.
In order to achieve good description of the backgrounds across all kinematics, only photons from kinematically matched events were combined using the method described in Appendix A. 
More detailed description of the improved event-mixing technique can be found in Ref.~\cite{PhD:Mineeva}. 
The number of $\pi^{0}$’s was calculated  from the integral of the Gaussian function situated on top of 4th-order polynomial of the event-mixed background.

\begin{figure}
  \centering
\includegraphics[width=0.49\textwidth]{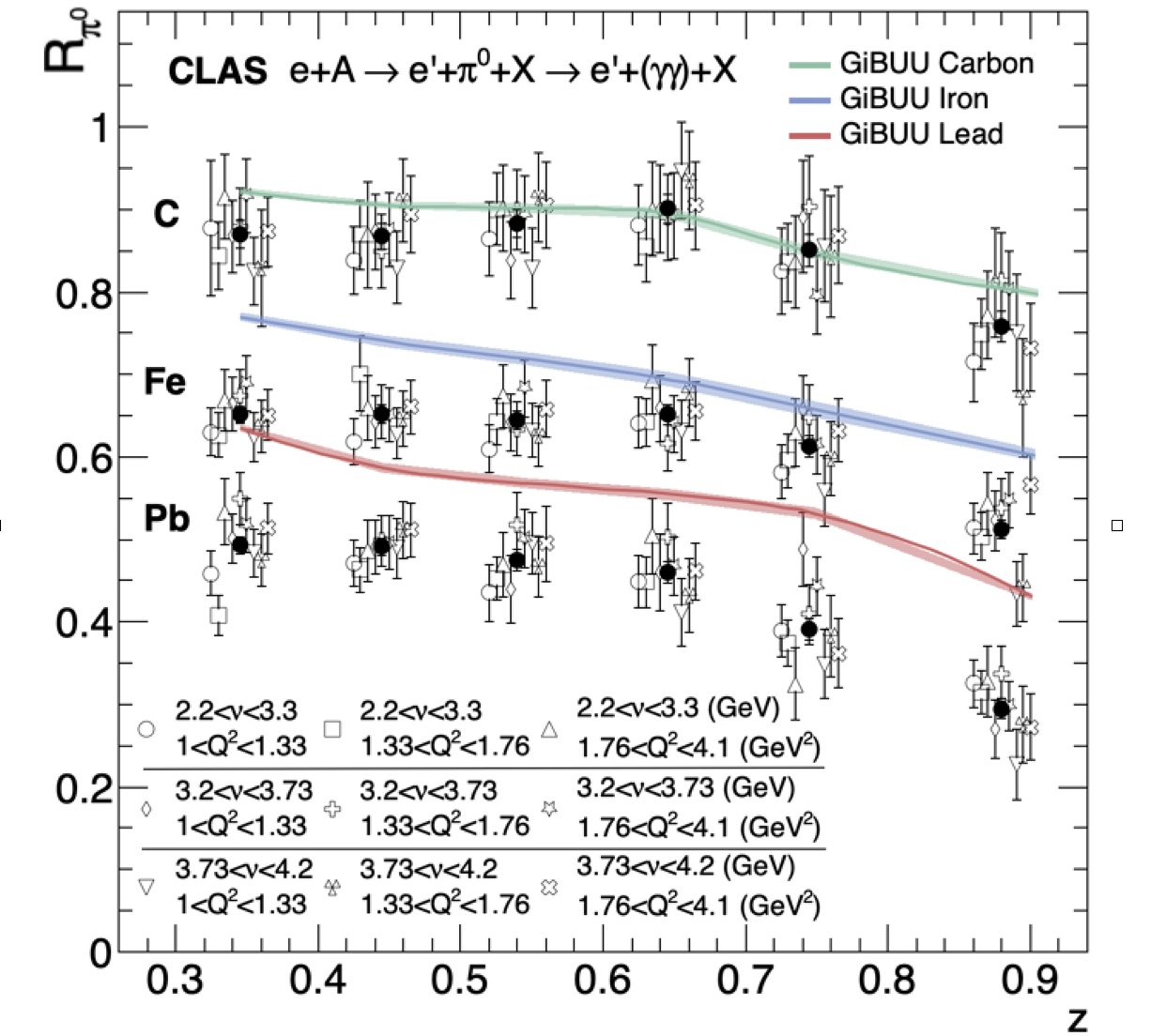}
  \caption{\small Three-dimensional $\pi^{0}$ multiplicity ratios for C, Fe, and Pb in ($Q^{2}$,~$\nu$,~$z$) bins plotted as a function of $z$. Each one of the six bins in $z$ contains 9 data points that correspond to the combination of the 3 bins of $\nu$ and 3 bins in $Q^{2}$. Each of the 9 points in $z$ is shifted around the center value of the bin; the points are plotted together with its statistical and systematic uncertainties.  The full circles correspond to the average value of multiplicity ratio in a given $z$ bin. The solid bands represent GiBUU model for each target with the width of the band indicating statistical errors of the simulations. }
\label{fig:Rvszqv}
\end{figure}
 \begin{figure*}[h!tb]
\begin{center}
 \includegraphics[width=\linewidth]{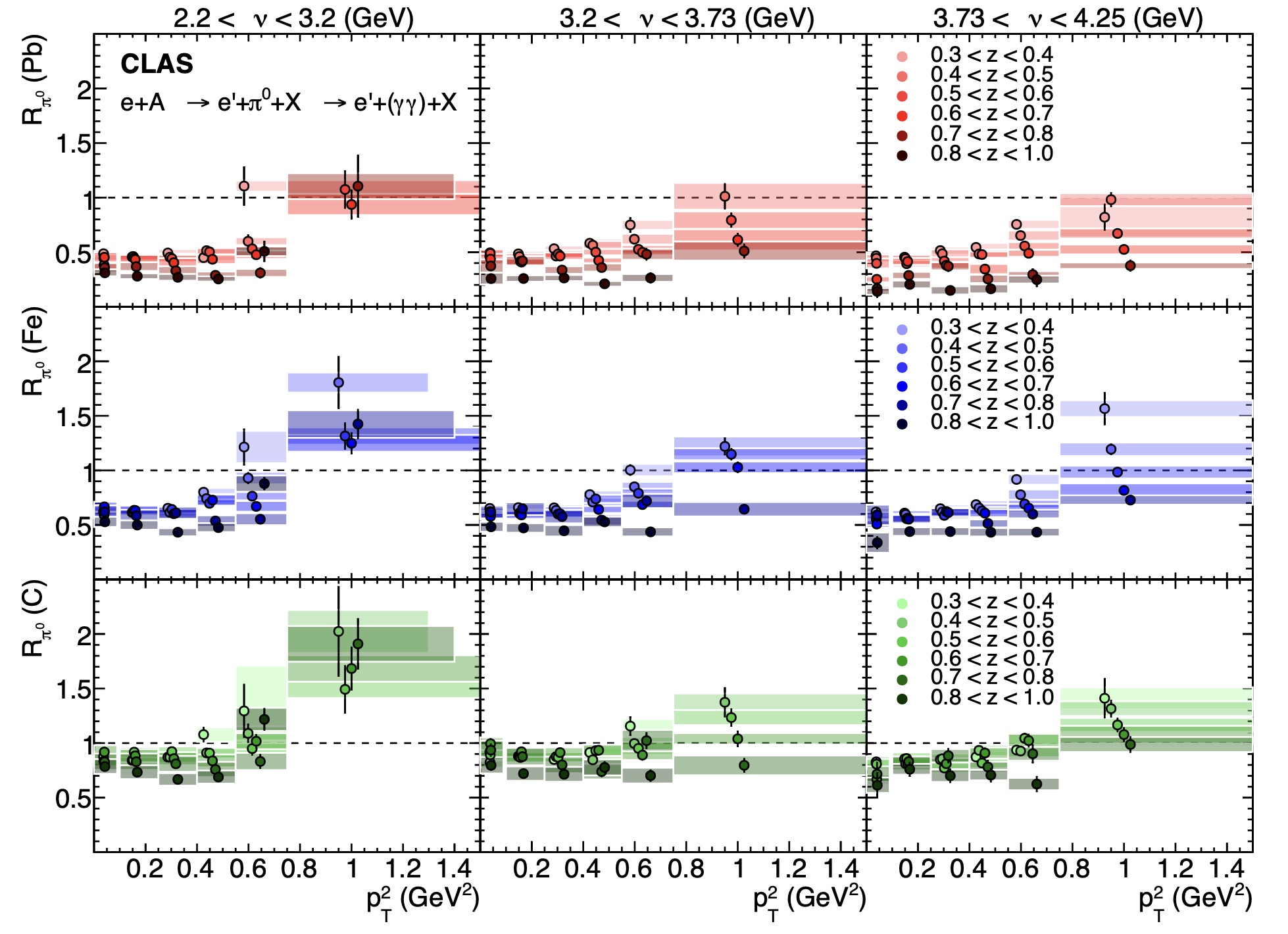}
 \caption{\small $\pi^{0}$ multiplicity ratios for C, Fe, and Pb in  ($\nu$,~$z$,~$p_{T}^{2}$) bins plotted as a function of $p_{T}^{2}$ in bins of $\nu$ (top horizontal line) and $z$  (indicated by the color). Points are shifted for ease of visualization around the mean value of $p_{T}^{2}$. Statistical uncertainties are indicated by black vertical lines; systematic uncertainties by the color bars. Horizontal uncertainties are related to the size of the bin: while for most bins in $p_{T}^{2}$ they are the same for each bin in $z$ and target, a few bins have smaller uncertainty bands related to the interval of data significance in the bin.}
 \label{fig:RvzpT}
\end{center}
\end{figure*}
\begin{figure*}[h!tb]
\begin{center}
 \includegraphics[width=\linewidth]{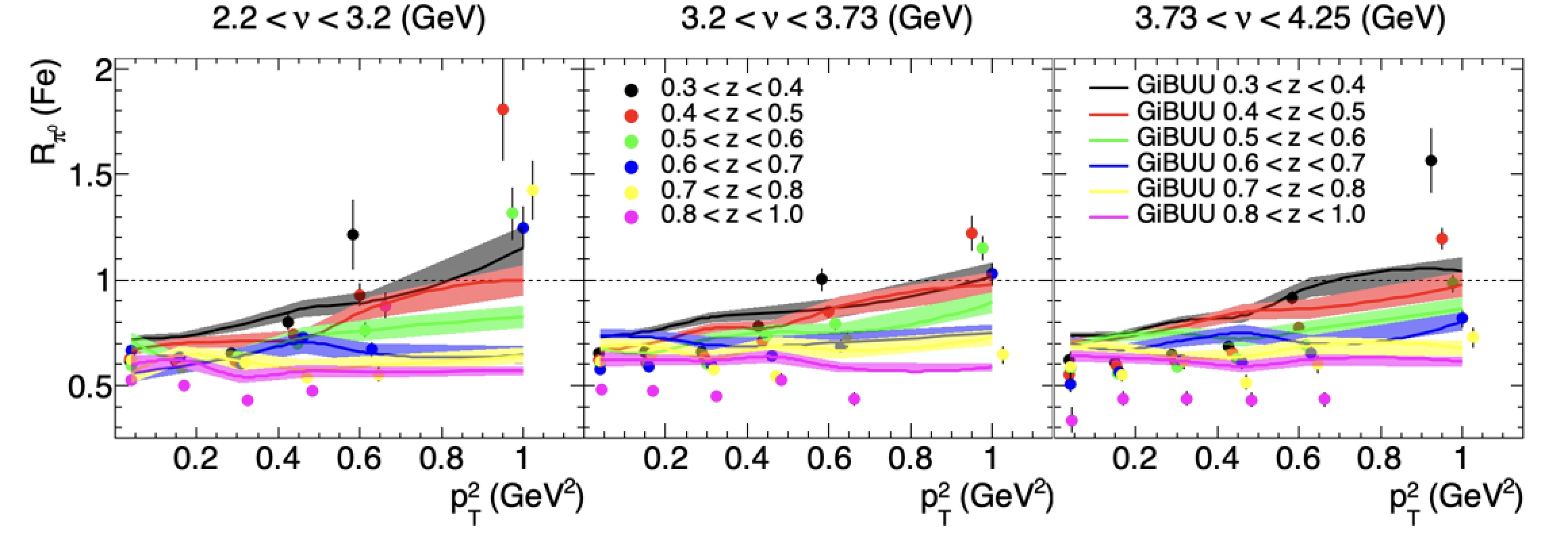}
 \caption{\small $\pi^{0}$ multiplicity ratios for Fe in  ($\nu$,~$z$,~$p_{T}^{2}$) as a function of $p_{T}^{2}$ in bins of $\nu$ (each panel) and $z$ (indicated by the color). Markers correspond to data points, the solid lines correspond to GiBUU projections with the width of the band indicating statistical errors of the simulation.}
 \label{fig:gibuu}
\end{center}
\end{figure*}


\section{III. CORRECTIONS}
The multiplicity ratio of Eq.~\ref{eq:R} can be viewed as a $\pi^{0}$ number ratio $N^{A/D}_{\pi^{0}}$ normalized by the electron number ratio $N^{D/A}_{e}$. Corrections to $N^{D/A}_{e}$ include:
(i) acceptance correction factors: these range from unity up to 8\%; 
(ii) radiative corrections due to internal radiation associated with bremsstrahlung off the nucleon: increase the multiplicity ratio up to 3\%;
(iii) radiative corrections due to Coulomb distortion in the field of the nucleus: decrease the multiplicity ratio down to 4\% with the largest corrections for Pb. 
Internal radiative corrections were calculated based on the Mo and Tsai formalism~\cite{RevModPhys.41.205} while Coulomb corrections - on the effective momentum approximation~\cite{ASTE2005}. Both are incorporated in the \textsc{EXTERNAL} code~\cite{dasu1994measurement}. The CLAS detector response was simulated with the GSIM package, based on GEANT3, which includes the locations and materials of the dual-target. 
The external radiative corrections that are associated with bremsstrahlung in the target material were incorporated in the GEANT3 simulations, and were accounted for by applying acceptance correction factors.
Corrections applied to $N^{A/D}_{\pi^{0}}$ for SIDIS $\pi^{0}$s include:
(i) acceptance correction factors: these change the multiplicity ratio
from -17\% to +8\% for ($\nu$,~$z$,~$p_{T}^{2}$) bins and 
from -14\% to +4\% for ($Q^{2}$,~$\nu$,~$z$);
(ii) radiative corrections for SIDIS $\pi^{0}$s: these affect the multiplicity ratio by less than 0.5\%. The latter were calculated using the HAPRAD code~\cite{refId0} that was modified using empirically derived nuclear structure functions. 
The combined effect of $N^{D/A}_{e}$ and $N^{A/D}_{\pi^{0}}$ radiative corrections does not exceed 4.8\%. 
Finally, due to the presence of the 15~$\mu$m aluminum walls (endcaps) of the liquid-deuterium target cell, we corrected multiplicity ratio for the $N^{D}_e$ and $N^{D}_{\pi^{0}}$ resulting in less than 1$\%$ correction.
Acceptance correction factors were calculated on a bin-by-bin basis using the LEPTO event generator 6.5.1~\cite{INGELMAN1997108}, modified to include nuclear Fermi motion of the target nucleon according to the Ciofi-Simula parametrization~\cite{PhysRevC.53.1689}.

Systematic uncertainties of the measurement are comprised of the following: (i) electron identification: target selection cuts, 
EC sampling fraction cuts, $\pi^-$ contamination, DC fiducial cuts, and electron radiative corrections;
(ii) photon identification:  cut on minimum energy deposited in EC,
time cut $\Delta t$, EC fiducial cuts;
and (iii) $\pi^0$ identification: background and signal shapes of the invariant mass distribution, acceptance corrections, and SIDIS radiative corrections. Systematic uncertainties were evaluated independently for each set of bins, ($\nu$,~$z$,~$p_{T}^{2}$) or ($Q^{2}$,~$\nu$,~$z$), for each nuclear ratio and applied either as a normalization or as a bin-by-bin uncertainty.
The largest contribution to the normalization-type uncertainty came from target vertex identification.
It results in 3.1$\%$, 2.4$\%$ and 2.3$\%$, for C, Fe and 
and Pb, respectively, in the ($\nu$,~$z$,~$p_{T}^{2}$) set of bins, and slightly smaller values for the ($Q^{2}$,~$\nu$,~$z$) bins.
The dominant source of the bin-by-bin systematic uncertainty is the $\pi^{0}$ invariant mass fit.
This uncertainty included both uncertainties on the background and signal shapes ranging on average from 1.4$\%$ for Fe in ($Q^{2}$,~$\nu$,~$z$) bins to 4.7$\%$ for Pb in ($\nu$,~$z$,~$p_{T}^{2}$) bins.
The total average systematic uncertainties in ($Q^{2}$,~$\nu$,~$z$) are 5.0$\%$, 4.9$\%$ and 6.9$\%$ for C, Fe and Pb multiplicities correspondingly;
in ($\nu$,~$z$,~$p_{T}^{2}$) they average to 7.1$\%$, 7.1$\%$ and 9.6$\%$ for C, Fe and Pb, respectively.
The average statistical uncertainty is typically several percent less.

The data supporting this study's results are available in the supplemental material to this article.

\section{IV. RESULTS AND DISCUSSION}
The three-fold differential $\pi^{0}$ multiplicity ratios are presented in bins of ($Q^{2}$,~$\nu$,~$z$) integrated over $p_{T}^{2}$ (Fig.~\ref{fig:Rvszqv}) and in bins of ($\nu$,~$z$,~$p_{T}^{2}$) integrated over $Q^{2}$ (Fig.~\ref{fig:RvzpT}). 
We compare our results (Fig.~\ref{fig:Rvszqv} and Fig.~\ref{fig:gibuu}) with those obtained from calculations using the GiBUU Monte Carlo program~\cite{Buss_2012} using default parametrization of GiBUU 2019 and employing the same kinematic selections as in our data.

In Fig.~\ref{fig:Rvszqv}, the multiplicity ratios are shown as a function of $z$ in bins of ($Q^{2}$, $\nu$). The data points are plotted alongside the model predictions. The data exhibit a flat behavior in the range $0.3 < z < 0.65$ and a monotonic decrease for higher values of $z$.
There is a notable dependence on nuclear size, indicating a path length-dependent process. For the smallest nucleus, carbon, suppression ranges from approximately 10\% at moderate $z$ to about 25\% at the highest $z$. In contrast, for the largest nucleus, lead, suppression varies from 50\% at lower $z$ to approximately 75\% at the highest $z$.

In modern energy loss models~\cite{Majumder:2010qh}, this attenuation is attributed to the assumption that a propagating quark emits multiple gluons as it traverses the nuclear medium; the larger the nucleus, the greater the gluon emission and the associated quark energy losses.
Within the framework of the GiBUU (Giessen Boltzmann-Uehling-Uhlenbeck) transport model~\cite{Gallmeister_2008}, which is largely based on elastic and inelastic pre-hadronic final-state interactions, the overall attenuation is understood in terms of pure hadron absorption resulting from an increased interaction time with the nuclear medium.
The data are qualitatively described by GiBUU over the entire $z$ range for all targets. However, nuclear attenuation is underestimated for larger nuclei. This behavior resembles the trends observed in charged pion and K$^{0}$ multiplicity ratio measurements from the same run period~\cite{CLAS:2021jhm, Daniel:2011nq}.

From Fig.~\ref{fig:Rvszqv}, no significant dependence on energy and momentum transfer to the system, \textit{i.e.}, $Q^{2}$ and $\nu$, is observed. However, the range of CLAS kinematics is much narrower than that of HERMES, where such dependencies were observed.

In Figure~\ref{fig:RvzpT}, multiplicity ratios are shown as a function of $p_{T}^{2}$ in bins of ($\nu$, $z$). The overall trend for all three targets is an enhancement of $R_{\pi^{0}}$ at high $p_{T}^{2}$ and low $z$, along with an overall atenuation as $z$ increases.
$R_{\pi^{0}}$ exhibits a pronounced dependence on $p_{T}^{2}$ in correlation with $z$. It remains independent of $p_{T}^{2}$ for all values of $z$ providing $p_{T}^{2}~<~0.5$~GeV$^2$. However, it increases rapidly for large $p_{T}^{2}$ and small $z$, reaching values exceeding unity.
The nuclear ordering of $R_{\pi^{0}}$ enhancement at high $p_{T}^{2}$ compared to low $p_{T}^{2}$ indicates that the relative enhancement is largest for heavier nuclei, such as lead and iron, while the smallest relative enhancement is observed for carbon. Additionally, a dependence on energy transfer $\nu$ is observed, with the enhancement being most pronounced at the lowest values of $\nu$.
Comparing enhancement of multiplicity ratios at highest values of  $p_{T}^{2}$ for $\pi^{0}$ with that of $\pi^{+}$, we observe a smaller enhancement for $\pi^{0}$ at the lowest values of $z$. In the case of carbon, this enhancement for $\pi^{+}$ would be nearly four times larger than that for $\pi^{0}$.

The pattern of $R_{\pi^{0}}$ enhancement at low $z$ and high $p_{T}^{2}$ is commonly referred to as the Cronin effect~\cite{Cronin:1974zm}. It was first observed in measurements by the European Muon Collaboration (EMC)~\cite{EuropeanMuon:1991jmx}, the Fermilab E665 experiment~\cite{E665:1995utr}, and later confirmed by HERMES~\cite{HERMES:2011qjb} and CLAS~\cite{CLAS:2021jhm}. The $R_{\pi^{0}}$ behavior reported in this paper is qualitatively similar to the previous HERMES and CLAS measurements of charged pions.
The GiBUU model provides a good description of the Cronin enhancement, as shown in Fig.~\ref{fig:gibuu} for the case of iron. Multiplicity ratios are observed to exceed unity at the lowest values of $z$ and the highest values of $p_{T}^{2}$. However, GiBUU predicts a smaller enhancement at low $z$ compared to the data, suggesting that some aspects of the theoretical description at high $p_{T}^{2}$ and low to mid $z$ values are incomplete. GiBUU accounts for the Cronin effect through the combined contributions of hadronic multiple scattering, final-state interactions, formation times, and nuclear medium effects such as color transparency and shadowing.

In the limit $z\rightarrow$~1, the lifetime of the propagating quark vanishes, as it is not allowed to lose any energy and, therefore, cannot accumulate transverse momentum through re-scattering. In contrast, the low $z$ regime corresponds to the opposite behavior, leading to the enhancement of transverse momenta. This scenario also suggests that the attenuation in the limit $z\rightarrow$~1 is purely due to hadron absorption.
The dependence of the Cronin effect on nuclear size points to a competition between partonic multiple scattering, which causes enhancement, and hadronic inelastic scattering, which leads to suppression. \\
The data points that support the findings of this article are published below in Appendix B: Tabulated Multiplicity Ratio Results.

\section{V. SUMMARY AND CONCLUSIONS}
In summary, we present the first three-fold differential $\pi^{0}$ multiplicity ratio measurements in SIDIS off D, C, Fe, and Pb. The results are reported in two sets of bins: $R_{\pi^{0}}$($Q^{2}$, $\nu$, $z$) and $R_{\pi^{0}}$($\nu$, $z$, $p_{T}^{2}$). From the first set of bins, we observe a strong dependence of the ratios on nuclear size, with the largest suppression observed for the highest atomic number $A$. This suppression remains constant in the moderate $z$ range and decreases rapidly for leading hadrons ($z >$ 0.65). The maximum suppression ranges from 25\% for carbon to 75\% for lead. Over the kinematic range of the CLAS experiment, the suppression of neutral pions shows no dependence on the energy and momentum transfer to the system, \textit{i.e.}, $\nu$ and $Q^{2}$.
The second set of bins shows that the multiplicity ratios are enhanced above unity for large $p_{T}^{2}$ and small $z$ (Cronin effect). The nuclear ordering of the Cronin effect reveals that the most significant enhancement at high $p_{T}^{2}$ occurs for the heaviest nuclei, such as iron and lead. Additionally, there is a noticeable dependence on $\nu$, with the largest Cronin effect observed at the lowest energy transfers.
The $z$-dependence of the multiplicity ratios, as well as the Cronin effect, is qualitatively described by the GiBUU transport model. Overall, the observed suppression ($R_{\pi^{0}}$ < 1) and enhancement ($R_{\pi^{0}}$ > 1) patterns are consistent with previous measurements of the $\pi^{0}$'s charged isospin partners, $\pi^{+}$ and $\pi^{-}$, with the exception that the high $p_{T}^{2}$ enhancement is smaller for $\pi^{0}$ compared to the charged pions.

These measurements have been successfully extended with an 11~GeV electron beam in the Jefferson Lab experiment E12-06-117~\cite{CLAS12Proposal}.
Offering a wider range in $Q^{2}$ and $\nu$ and higher luminosity, a wealth of new opportunities is becoming feasible: access to the quark mass dependence of the hadronization with GeV-scale meson formation, extraction of four-fold multiplicities for a large spectrum of hadrons, and searches for diquark correlations in baryon formation~\cite{BARABANOV, PhysRevLett.130.142301}.
With its collider energies and extended range of kinematic variables, the proposed $eA$ program at the Electron-Ion Collider~\cite{AbdulKhalek:2021gbh} will provide new insights into hadronization mechanisms. This includes clean measurements of medium-induced energy loss in scenarios where hadrons are formed outside the nuclear medium, as well as studies of potentially very different hadronization properties for heavy mesons.

\subsection{ACKNOWLEDGEMENTS}
\begin{acknowledgments}
The authors would like to thank Dave Gaskell for fruitful discussions on radiative corrections with EXTERNAL code.
We acknowledge the staff of the Accelerator and the Physics Divisions at Jefferson Lab in making this experiment possible. 
This work is supported by the Chilean Agencia Nacional de Investigacion y Desarollo (ANID), FONDECYT grants No.1221827 and No.1181215, No.1161642 and No.1240904, ANID PIA/APOYO AFB230003, and by the ANID-Millennium Science Initiative Program - ICN2019\_044.
This work was supported in part by the U.S. Department of Energy (DOE) and National Science Foundation (NSF), the Italian Istituto Nazionale di Fisica Nucleare (INFN), the French Centre National de la Recherche Scientifique (CNRS), the French Commissariat á l’Energie Atomique (CEA), the Skobeltsyn Institute of Nuclear Physics (SINP), the Scottish Universities Physics Alliance (SUPA), the National Research Foundation of Korea (NRF), the UK Science and Technology Facilities Council (STFC). 
The Southeastern Universities Research Association (SURA) operates the Thomas Jefferson National
Accelerator Facility for the U.S. Department of Energy
under Contract No. DE-AC05-06OR23177.

\end{acknowledgments}

\bibliographystyle{apsrev4-1} 
\bibliography{PRL.bib}

\newpage
\onecolumngrid

\subsection{\label{sec:appendixA}Appendix A: Event Mixing for Background Subtraction}

\hspace{0.1cm} Due to the shortness of the $\pi^{0}$ lifetime, its direct detection at these energies is impossible,
therefore an invariant-mass analysis is the only procedure available when considering the SIDIS reaction. 
Construction of all possible combinations of two photons within one event produces a 
Gaussian-type peak distributed around the m$_{\pi^{0}}$ on top of the background. 
 However, since our data set is finely binned, and each bin is limited in statistics, the free-parameter polynomial fit of the background would lead to unnecessarily large uncertainties on the extracted number of $\pi^{0}$'s. 
 An \textit{a priori} knowledge of the background shape allows for a reduction in this uncertainty by adapting a polynomial fit to the background shape, 
 thus allowing to have only one free parameter: the normalization of the fit function.
 
In order to estimate the shape under the combinatorial background, a commonly used approach is that of  event mixing. 
It consists of constructing the invariant mass distribution by 
combining a single photon $\gamma_1$ from one event with a single photon $\gamma_2$ from another event into a pair m$_{\gamma_1\gamma_2}$. 
Such combinations are constructed over the entire data set.
Since our data is binned in ($Q^{2}$,~$\nu$,~$z$) and ($\nu$,~$z$,~$p_{T}^{2}$) bins, we divide mixed events distribution into those bins, fitting each bin with a fourth-order polynomial function. 
This allows for the pre-determination of coefficients for the polynomial function used to fit the background data.
This function will only have one free aforementioned parameter (corresponding to the parameter $p_0$ on the left of Fig.~{\color{blue}1)}. 

In general, such an approach works well to describe the background shape on a large kinematical range, however, in our case of multiple fine binning it fails to properly mimic the shape of the background across all the bins.
 The main reason why it fails is related to the difference in kinematics of the events used to construct a mixed pair. 
Combining two uncorrelated photons, that come from two different scattering events, can result in a mixed pair that carries neither direction nor the kinematics of an actual event. 
In other words, a mixed pair constructed from two uncorrelated events may violate kinematics with respect to the
 single-event pair of photons. 
 On a large kinematical range such differences average. 
 However, when the background shape is non-linear and the phase-space is restricted, such recombination will not respect the distribution of the actual two-photon invariant mass background.

This difference can be corrected by accounting for the global characteristics of mixed events.
To do this, we first restrict the phase-space available for choosing an uncorrelated photon:
the kinematics of each event from which $\gamma_1$ and $\gamma_2$ originate must be such that 
the scattered electron e$^\prime$ and one of the photons from the $\pi^{0}$ candidate from the event 1 must belong to the
same kinematical bin in ($Q^{2}$, $\nu$) and ($z$, $p_{T}^{2}$ ) as those from the event 2.
Next,  in order to mimic initial kinematical conditions as if two uncorrelated photons, $\gamma_1$ and $\gamma_2$, were originating from the same scattering event, we rotate one of the photons around the electron beam by the angular difference $\delta\phi$ between virtual photons $\gamma_1^*$ and $\gamma_2^*$. 
In other words, in order to match the initial kinematics of uncorrelated events, one of the events is rotated by the angle $\delta\phi$ so that the virtual photons of two uncorrelated events have the same $\phi$. 
One may also consider to rotate the entire event by $\delta\theta$ angle,
 such that the virtual photons from two uncorrelated events also have the same $\theta$. Such rotation, however,
  will shift the actual beam position in the rotated event, which is undesirable. Therefore, only the $\delta\phi$ rotation is performed.

The improvement of general event mixing proved accurate on the large sample of finely binned MC events with tagged photons. 
Providing a larger statistical sample of real data events, this special event mixing method could have been further refined, by, for example, imposing kinematical constraints on the difference between momentum transferred by the virtual photons between two events or on the kinematical characteristics of the resulting mixed m$_{\gamma_1\gamma_2}$. To conclude, \textit{a priori} knowledge of the background shape is beneficial as it provides a more accurate fit and, therefore, smaller uncertainty on the extracted N$_{\pi^{0}}$ as compared to the free-parameter polynomial, in particular in those bins where the background shape is rapidly changing, or in those where the overall statistics is small. 

\subsection{\label{sec:appendixB}Appendix B: Tabulated Multiplicity Ratio Results}
Table~\ref{tab:RvzpT} and Table~\ref{tab:Rqvz} tabulate numerical results of neutral pion multiplicity ratios, as illustrated in Figs.~{\color{blue}2} and {\color{blue}3}, respectively, with statistical and systematic uncertainties in each of the three-fold bins in ($Q^{2}$,~$\nu$,~$z$) and ($\nu$,~$z$,~$p_{T}^{2}$) for C, Fe and Pb nuclei. Multiplicity ratios are reported as: Multiplicity Ratio $\pm$ Statistical $\pm$  Systematical Uncertainties. 



\begin{table}[h!]
\centering
\caption{\label{tab:RvzpT} Numerical results for $\pi^{0}$ multiplicity ratios for C, Fe and Pb in ($\nu$,~$z$,~$p_{T}^{2}$) bins as depicted in Fig.~{\color{blue}3}.}
\centering
\hskip-1.0cm \begin{tabular}{c | c  c  c}
\hline \hline
\multirow{2}{*}{$p_{T}^{2}$-bin} & &  2.2 $< \nu <$ 3.2, 0.3 $< z <$ 0.4 &  \\ \cline{2-4}
 & Carbon & Iron & Lead\\ \hline
0. - 0.1 & 
 0.875	$\pm$	0.020	$\pm$	0.046	&	0.616	$\pm$	0.012	$\pm$	0.033	&	0.486	$\pm$	0.014	$\pm$	0.027 
 \\
0.1 - 0.25 &
0.841	$\pm$	0.015	$\pm$	0.044	&	0.614	$\pm$	0.010	$\pm$	0.032	&	0.456	$\pm$	0.011	$\pm$	0.027
\\
0.25 - 0.4 &
0.872	$\pm$	0.027	$\pm$	0.047	&	0.653	$\pm$	0.017	$\pm$	0.033	&	0.493	$\pm$	0.019	$\pm$	0.036
\\
0.4 - 0.55 &
1.078	$\pm$	0.072	$\pm$	0.066	&	0.800	$\pm$	0.041	$\pm$	0.039	&	0.45	$\pm$	0.043	$\pm$	0.029 \\ 
0.55 - 0.75 &
1.295	$\pm$	0.248	$\pm$	0.413	&	1.214	$\pm$	0.168	$\pm$	0.148	&	1.106	$\pm$	0.179	$\pm$	0.051
\\
0.75 - 1.5 &
-	&	-	&	-	
\\
\hline \hline
\multirow{2}{*}{$p_{T}^{2}$-bin} & &  3.2 $< \nu <$ 3.73, 0.3 $< z <$ 0.4 &  \\ \cline{2-4}
 & Carbon & Iron & Lead\\ \hline
0. - 0.1 & 
 0.917	$\pm$	0.021	$\pm$	0.047	&	0.652	$\pm$	0.013	$\pm$	0.033	&	0.464	$\pm$	0.014	$\pm$	0.025
 \\
0.1 - 0.25 &
0.879	$\pm$	0.014	$\pm$	0.045	&	0.659	$\pm$	0.009	$\pm$	0.032	&	0.485	$\pm$	0.010	$\pm$	0.026
\\
0.25 - 0.4 &
0.850	$\pm$	0.018	$\pm$	0.044	&	0.657	$\pm$	0.011	$\pm$	0.032	&	0.532	$\pm$	0.013	$\pm$	0.029
\\
0.4 - 0.55 &
0.914	$\pm$	0.032	$\pm$	0.047	&	0.778	$\pm$	0.022	$\pm$	0.038	&	0.582	$\pm$	0.025	$\pm$	0.043
\\
0.55 - 0.75 &
1.156	$\pm$	0.089	$\pm$	0.064	&	1.003	$\pm$	0.053	$\pm$	0.057	&	0.750	$\pm$	0.070	$\pm$	0.046
\\
0.75 - 1.5 &
-	&	-	&	-
\\
\hline \hline
\multirow{2}{*}{$p_{T}^{2}$-bin} & &  3.73 $< \nu <$ 4.25, 0.3 $< z <$ 0.4 &  \\ \cline{2-4}
 & Carbon & Iron & Lead\\ \hline

0. - 0.1 & 
 0.823	$\pm$	0.023	$\pm$	0.047	&	0.619	$\pm$	0.015	$\pm$	0.037	&	0.467	$\pm$	0.017	$\pm$	0.032
 \\

0.1 - 0.25 &
0.855	$\pm$	0.016	$\pm$	0.049	&	0.609	$\pm$	0.009	$\pm$	0.036	&	0.454	$\pm$	0.010	$\pm$	0.031
\\
0.25 - 0.4 &
0.849	$\pm$	0.018	$\pm$	0.048	&	0.649	$\pm$	0.011	$\pm$	0.038	&	0.515	$\pm$	0.013	$\pm$	0.036
\\
0.4 - 0.55 &
0.873	$\pm$	0.026	$\pm$	0.049	&	0.686	$\pm$	0.016	$\pm$	0.039	&	0.544	$\pm$	0.018	$\pm$	0.040
\\
0.55 - 0.75 &
0.937	$\pm$	0.042	$\pm$	0.070	&	0.917	$\pm$	0.033	$\pm$	0.049	&	0.755	$\pm$	0.038	$\pm$	0.042
\\

0.75 - 1.5 &
1.411	$\pm$	0.185	$\pm$	0.100	&	1.566	$\pm$	0.151	$\pm$	0.077	&	0.820	$\pm$	0.123	$\pm$	0.181
\\
\hline \hline
\multirow{2}{*}{$p_{T}^{2}$-bin} & &  2.2 $< \nu <$ 3.2, 0.4 $< z <$ 0.5 &  \\ \cline{2-4}
 & Carbon & Iron & Lead\\ \hline
0. - 0.1 & 
0.833	$\pm$	0.021	$\pm$	0.048	&	0.627	$\pm$	0.014	$\pm$	0.038	&	0.447	$\pm$	0.015	$\pm$	0.03
 \\
0.1 - 0.25 &
0.843	$\pm$	0.017	$\pm$	0.048	&	0.617	$\pm$	0.011	$\pm$	0.036	&	0.461	$\pm$	0.012	$\pm$	0.032
\\
0.25 - 0.4 &
0.871	$\pm$	0.023	$\pm$	0.049	&	0.616	$\pm$	0.014	$\pm$	0.036	&	0.454	$\pm$	0.015	$\pm$	0.033
\\
0.4 - 0.55 &
0.910	$\pm$	0.038	$\pm$	0.054	&	0.742	$\pm$	0.025	$\pm$	0.041	&	0.514	$\pm$	0.028	$\pm$	0.033
\\
0.55 - 0.75 &
1.089	$\pm$	0.088	$\pm$	0.064	&	0.930	$\pm$	0.055	$\pm$	0.057	&	0.6	$\pm$	0.062	$\pm$	0.037
\\
0.75 - 1.5 &
2.024	$\pm$	0.416	$\pm$	0.196	&	1.805	$\pm$	0.242	$\pm$	0.092	&	-	
\\ \hline \hline
\multirow{2}{*}{$p_{T}^{2}$-bin} & &  3.2 $< \nu <$ 3.73, 0.4 $< z <$ 0.5 &  \\ \cline{2-4}
 & Carbon & Iron & Lead\\ \hline
0. - 0.1 & 
0.823	$\pm$	0.026	$\pm$	0.048	&	0.623	$\pm$	0.017	$\pm$	0.036	&	0.494	$\pm$	0.020	$\pm$	0.037
\\
0.1 - 0.25 &
0.876	$\pm$	0.019	$\pm$	0.049	&	0.608	$\pm$	0.011	$\pm$	0.034	&	0.460	$\pm$	0.013	$\pm$	0.029
\\
0.25 - 0.4 &
0.876	$\pm$	0.021	$\pm$	0.049	&	0.634	$\pm$	0.012	$\pm$	0.035	&	0.465	$\pm$	0.014	$\pm$	0.032
\\
0.4 - 0.55 &
0.847	$\pm$	0.028	$\pm$	0.049	&	0.710	$\pm$	0.018	$\pm$	0.038	&	0.563	$\pm$	0.022	$\pm$	0.039
\\
0.55 - 0.75 &
0.996	$\pm$	0.045	$\pm$	0.059	&	0.850	$\pm$	0.029	$\pm$	0.044	&	0.619	$\pm$	0.034	$\pm$	0.049
\\
0.75 - 1.5 &
1.374	$\pm$	0.137	$\pm$	0.081	&	1.220	$\pm$	0.082	$\pm$	0.090	&	1.012	$\pm$	0.121	$\pm$	0.123
\\ \hline \hline
(Continued)
\end{tabular}
\end{table}

\begin{table}[h!]
\centering
\hskip-1.0cm \begin{tabular}{c | c  c  c}
\hline \hline
\multirow{2}{*}{$p_{T}^{2}$-bin} & &  3.73 $< \nu <$ 4.25, 0.4 $< z <$ 0.5 &  \\ \cline{2-4}
 & Carbon & Iron & Lead\\ \hline
0. - 0.1 & 
0.828	$\pm$	0.041	$\pm$	0.051	&	0.552	$\pm$	0.023	$\pm$	0.036	&	0.444	$\pm$	0.027	$\pm$	0.035
\\
0.1 - 0.25 &
0.811	$\pm$	0.023	$\pm$	0.047	&	0.596	$\pm$	0.014	$\pm$	0.037	&	0.441	$\pm$	0.015	$\pm$	0.035
\\
0.25 - 0.4 &
0.856	$\pm$	0.024	$\pm$	0.048	&	0.619	$\pm$	0.014	$\pm$	0.037	&	0.486	$\pm$	0.016	$\pm$	0.040
\\
0.4 - 0.55 &
0.934	$\pm$	0.032	$\pm$	0.051	&	0.655	$\pm$	0.018	$\pm$	0.039	&	0.484	$\pm$	0.021	$\pm$	0.041
\\
0.55 - 0.75 &
0.927	$\pm$	0.037	$\pm$	0.057	&	0.777	$\pm$	0.024	$\pm$	0.048	&	0.652	$\pm$	0.031	$\pm$	0.047
\\
0.75 - 1.5 &
1.316	$\pm$	0.081	$\pm$	0.068	&	1.193	$\pm$	0.052	$\pm$	0.062	&	0.981	$\pm$	0.069	$\pm$	0.059
\\ \hline \hline

\multirow{2}{*}{$p_{T}^{2}$-bin} & &  2.2 $< \nu <$ 3.2, 0.5 $< z <$ 0.6 &  \\ \cline{2-4}
 & Carbon & Iron & Lead\\ \hline
0. - 0.1 & 
0.834	$\pm$	0.026	$\pm$	0.044	&	0.593	$\pm$	0.016	$\pm$	0.033	&	0.381	$\pm$	0.017	$\pm$	0.038
\\
0.1 - 0.25 &
0.917	$\pm$	0.024	$\pm$	0.047	&	0.634	$\pm$	0.014	$\pm$	0.034	&	0.446	$\pm$	0.015	$\pm$	0.034
\\
0.25 - 0.4 &
0.922	$\pm$	0.027	$\pm$	0.047	&	0.642	$\pm$	0.016	$\pm$	0.036	&	0.442	$\pm$	0.017	$\pm$	0.033
\\
0.4 - 0.55 &
0.909	$\pm$	0.037	$\pm$	0.047	&	0.698	$\pm$	0.024	$\pm$	0.037	&	0.503	$\pm$	0.029	$\pm$	0.034
\\
0.55 - 0.75 &
0.949	$\pm$	0.063	$\pm$	0.050	&	0.763	$\pm$	0.037	$\pm$	0.039	&	0.531	$\pm$	0.048	$\pm$	0.033
\\
0.75 - 1.5 &
1.493	$\pm$	0.221	$\pm$	0.079	&	1.315	$\pm$	0.123	$\pm$	0.079	&	1.074	$\pm$	0.176	$\pm$	0.085
\\ \hline \hline
\multirow{2}{*}{$p_{T}^{2}$-bin} & &  3.2 $< \nu <$ 3.73, 0.5 $< z <$ 0.6 &  \\ \cline{2-4}
 & Carbon & Iron & Lead\\ \hline
0. - 0.1 & 
0.893	$\pm$	0.038	$\pm$	0.052	&	0.595	$\pm$	0.021	$\pm$	0.033	&	0.489	$\pm$	0.026	$\pm$	0.037
\\
0.1 - 0.25 &
0.868	$\pm$	0.026	$\pm$	0.046	&	0.600	$\pm$	0.015	$\pm$	0.033	&	0.416	$\pm$	0.016	$\pm$	0.029
\\
0.25 - 0.4 &
0.870	$\pm$	0.027	$\pm$	0.046	&	0.600	$\pm$	0.015	$\pm$	0.033	&	0.483	$\pm$	0.019	$\pm$	0.035
\\
0.4 - 0.55 &
0.931	$\pm$	0.036	$\pm$	0.049	&	0.739	$\pm$	0.022	$\pm$	0.038	&	0.501	$\pm$	0.024	$\pm$	0.029
\\
0.55 - 0.75 &
0.952	$\pm$	0.044	$\pm$	0.058	&	0.790	$\pm$	0.029	$\pm$	0.041	&	0.526	$\pm$	0.033	$\pm$	0.038
\\
0.75 - 1.5 &
1.234	$\pm$	0.083	$\pm$	0.069	&	1.148	$\pm$	0.057	$\pm$	0.055	&	0.794	$\pm$	0.068	$\pm$	0.080
\\ \hline \hline
\multirow{2}{*}{$p_{T}^{2}$-bin} & &  2.2 $< \nu <$ 3.2, 0.5 $< z <$ 0.6 &  \\ \cline{2-4}
 & Carbon & Iron & Lead\\ \hline
0. - 0.1 & 
0.807	$\pm$	0.065	$\pm$	0.047	&	0.584	$\pm$	0.038	$\pm$	0.033	&	0.397	$\pm$	0.041	$\pm$	0.045
\\
0.1 - 0.25 &
0.859	$\pm$	0.034	$\pm$	0.056	&	0.557	$\pm$	0.018	$\pm$	0.032	&	0.394	$\pm$	0.021	$\pm$	0.030
\\
0.25 - 0.4 &
0.772	$\pm$	0.030	$\pm$	0.046	&	0.587	$\pm$	0.019	$\pm$	0.034	&	0.419	$\pm$	0.021	$\pm$	0.035
\\
0.4 - 0.55 &
0.817	$\pm$	0.035	$\pm$	0.047	&	0.629	$\pm$	0.023	$\pm$	0.035	&	0.478	$\pm$	0.027	$\pm$	0.029
\\
0.55 - 0.75 &
1.044	$\pm$	0.051	$\pm$	0.057	&	0.692	$\pm$	0.027	$\pm$	0.037	&	0.557	$\pm$	0.035	$\pm$	0.035
\\
0.75 - 1.5 &
1.167	$\pm$	0.067	$\pm$	0.064	&	0.984	$\pm$	0.041	$\pm$	0.059	&	0.672	$\pm$	0.048	$\pm$	0.046
\\ \hline \hline

\hline \hline
\multirow{2}{*}{$p_{T}^{2}$-bin} & &  2.2 $< \nu <$ 3.2, 0.6 $< z <$ 0.7 &  \\ \cline{2-4}
 & Carbon & Iron & Lead\\ \hline
0. - 0.1 & 
0.918	$\pm$	0.035	$\pm$	0.062	&	0.666	$\pm$	0.021	$\pm$	0.046	&	0.454	$\pm$	0.023	$\pm$	0.046
\\
0.1 - 0.25 &
0.884	$\pm$	0.029	$\pm$	0.06	&	0.633	$\pm$	0.018	$\pm$	0.045	&	0.433	$\pm$	0.02	$\pm$	0.042
\\
0.25 - 0.4 &
0.858	$\pm$	0.031	$\pm$	0.059	&	0.603	$\pm$	0.018	$\pm$	0.043	&	0.404	$\pm$	0.02	$\pm$	0.04
\\
0.4 - 0.55 &
0.838	$\pm$	0.041	$\pm$	0.058	&	0.728	$\pm$	0.028	$\pm$	0.05	&	0.435	$\pm$	0.032	$\pm$	0.044
\\
0.55 - 0.75 &
1.017	$\pm$	0.066	$\pm$	0.073	&	0.67	$\pm$	0.034	$\pm$	0.049	&	0.476	$\pm$	0.046	$\pm$	0.038
\\
0.75 - 1.5 &
1.684	$\pm$	0.2	$\pm$ 0.121	&	1.248	$\pm$	0.099	$\pm$	0.077	&	0.937	$\pm$	0.136	$\pm$	0.097
\\ \hline \hline
\multirow{2}{*}{$p_{T}^{2}$-bin} & &  3.2 $< \nu <$ 3.73, 0.6 $< z <$ 0.7 &  \\ \cline{2-4}
 & Carbon & Iron & Lead\\ \hline
0. - 0.1 & 
0.994	$\pm$	0.051	$\pm$	0.058	&	0.578	$\pm$	0.026	$\pm$	0.032	&	0.437	$\pm$	0.032	$\pm$	0.048
\\
0.1 - 0.25 &
0.921	$\pm$	0.036	$\pm$	0.047	&	0.591	$\pm$	0.019	$\pm$	0.036	&	0.409	$\pm$	0.023	$\pm$	0.031
\\
0.25 - 0.4 &
0.913	$\pm$	0.035	$\pm$	0.048	&	0.602	$\pm$	0.02	$\pm$	0.034	&	0.464	$\pm$	0.025	$\pm$	0.033
\\
0.4 - 0.55 &
0.935	$\pm$	0.047	$\pm$	0.053	&	0.641	$\pm$	0.024	$\pm$	0.032	&	0.424	$\pm$	0.03	$\pm$	0.025
\\
0.55 - 0.75 &
0.89	$\pm$	0.05	$\pm$	0.045	&	0.687	$\pm$	0.029	$\pm$	0.034	&	0.498	$\pm$	0.039	$\pm$	0.026
\\
0.75 - 1.5 &
1.039	$\pm$	0.074	$\pm$	0.052	&	1.028	$\pm$	0.053	$\pm$	0.053	&	0.613	$\pm$	0.063	$\pm$	0.098
\\ \hline \hline
\multirow{2}{*}{$p_{T}^{2}$-bin} & &  3.73 $< \nu <$ 4.2, 0.6 $< z <$ 0.7 &  \\ \cline{2-4}
 & Carbon & Iron & Lead\\ \hline
0. - 0.1 &
0.668	$\pm$	0.061	$\pm$	0.044	&	0.508	$\pm$	0.038	$\pm$	0.032	&	0.251	$\pm$	0.041	$\pm$	0.027
\\
0.1 - 0.25 &
0.813	$\pm$	0.044	$\pm$	0.061	&	0.563	$\pm$	0.025	$\pm$	0.040	&	0.415	$\pm$	0.029	$\pm$	0.031
\\
0.25 - 0.4 &
0.811	$\pm$	0.041	$\pm$	0.051	&	0.623	$\pm$	0.026	$\pm$	0.040	&	0.380	$\pm$	0.027	$\pm$	0.032
\\
0.4 - 0.55 &
0.909	$\pm$	0.052	$\pm$	0.052	&	0.608	$\pm$	0.028	$\pm$	0.036	&	0.345	$\pm$	0.034	$\pm$	0.047
\\
0.55 - 0.75 &
1.023	$\pm$	0.063	$\pm$	0.063	&	0.654	$\pm$	0.033	$\pm$	0.038	&	0.490	$\pm$	0.045	$\pm$	0.033
\\
0.75 - 1.5 &
1.078	$\pm$	0.067	$\pm$	0.081	&	0.816	$\pm$	0.039	$\pm$	0.065	&	0.525	$\pm$	0.045	$\pm$	0.045
\\ \hline \hline
\multirow{2}{*}{$p_{T}^{2}$-bin} & &  2.2 $< \nu <$ 3.2, 0.7 $< z <$ 0.8 &  \\ \cline{2-4}
 & Carbon & Iron & Lead\\ \hline
0. - 0.1 & 
0.828	$\pm$	0.038	$\pm$	0.075	&	0.620	$\pm$	0.023	$\pm$	0.059	&	0.358	$\pm$	0.024	$\pm$	0.042
\\
0.1 - 0.25 &
0.830	$\pm$	0.034	$\pm$	0.075	&	0.584	$\pm$	0.020	$\pm$	0.055	&	0.368	$\pm$	0.023	$\pm$	0.040
\\
0.25 - 0.4 &
0.811	$\pm$	0.037	$\pm$	0.073	&	0.611	$\pm$	0.022	$\pm$	0.056	&	0.330	$\pm$	0.026	$\pm$	0.041
\\
0.4 - 0.55 &
0.759	$\pm$	0.047	$\pm$	0.070	&	0.536	$\pm$	0.026	$\pm$	0.051	&	0.288	$\pm$	0.033	$\pm$	0.035
\\
0.55 - 0.75 &
0.832	$\pm$ 0.067	 $\pm$	0.079	&	0.552	$\pm$	0.035	$\pm$	0.055	&	0.310	$\pm$	0.050	$\pm$	0.037
\\
0.75 - 1.5 &
1.910	$\pm$	0.235	$\pm$	0.164	&	1.425	$\pm$	0.137	$\pm$	0.126	&	1.105	$\pm$	0.289	$\pm$	0.119
\\ \hline \hline
(Continued)
\end{tabular}
\end{table}

\begin{table}[h]
\centering
\hskip-1.0cm \begin{tabular}{c | c  c  c}
\hline \hline
\multirow{2}{*}{$p_{T}^{2}$-bin} & &  3.2 $< \nu <$ 3.73, 0.7 $< z <$ 0.8 &  \\ \cline{2-4}
 & Carbon & Iron & Lead\\ \hline
0. - 0.1 & 
0.935	$\pm$	0.062	$\pm$	0.087	&	0.618	$\pm$	0.033	$\pm$	0.058	&	0.374	$\pm$	0.040	$\pm$	0.060
\\
0.1 - 0.25 &
0.876	$\pm$	0.047	$\pm$	0.080	&	0.648	$\pm$	0.026	$\pm$	0.059	&	0.421	$\pm$	0.033	$\pm$	0.043
\\
0.25 - 0.4 &
0.801	$\pm$	0.042	$\pm$	0.076	&	0.577	$\pm$	0.025	$\pm$	0.062	&	0.336	$\pm$	0.029	$\pm$	0.035
\\
0.4 - 0.55 &
0.739	$\pm$	0.049	$\pm$	0.071	&	0.544	$\pm$	0.028	$\pm$	0.051	&	0.359	$\pm$	0.037	$\pm$	0.038
\\
0.55 - 0.75 &
1.022	$\pm$	0.077	$\pm$	0.098	&	0.719	$\pm$	0.041	$\pm$	0.065	&	0.481	$\pm$	0.058	$\pm$	0.051
\\
0.75 - 1.5 &
0.796	$\pm$	0.066	$\pm$	0.090	&	0.644	$\pm$	0.040	$\pm$	0.066	&	0.510	$\pm$	0.066	$\pm$	0.088
\\ \hline \hline

\multirow{2}{*}{$p_{T}^{2}$-bin} & &  3.73 $< \nu <$ 4.2, 0.7 $< z <$ 0.8 &  \\ \cline{2-4}
 & Carbon & Iron & Lead\\ \hline
0. - 0.1 & 
0.716	$\pm$	0.112	$\pm$	0.071	&	0.592	$\pm$	0.069	$\pm$	0.094	&	0.165	$\pm$	0.060	$\pm$	0.025
\\
0.1 - 0.25 &
0.831	$\pm$	0.064	$\pm$	0.049	&	0.553	$\pm$	0.035	$\pm$	0.039	&	0.287	$\pm$	0.044	$\pm$	0.032
\\
0.25 - 0.4 &
0.885	$\pm$	0.069	$\pm$	0.061	&	0.613	$\pm$	0.037	$\pm$	0.040	&	0.370	$\pm$	0.050	$\pm$	0.036
\\
0.4 - 0.55 &
0.781	$\pm$	0.066	$\pm$	0.048	&	0.514	$\pm$	0.035	$\pm$	0.037	&	0.255	$\pm$	0.052	$\pm$	0.038
\\
0.55 - 0.75 &
0.902	$\pm$	0.086	$\pm$	0.063	&	0.600	$\pm$	0.042	$\pm$	0.039	&	0.294	$\pm$	0.057	$\pm$	0.031
\\
0.75 - 1.5 &
0.988	$\pm$	0.076	$\pm$	0.069	&	0.728	$\pm$	0.046	$\pm$	0.043	&	0.376	$\pm$	0.055	$\pm$	0.030
\\ \hline \hline
\multirow{2}{*}{$p_{T}^{2}$-bin} & &  2.2 $< \nu <$ 3.2, 0.8 $< z <$ 1.0 &  \\ \cline{2-4}
 & Carbon & Iron & Lead\\ \hline
0. - 0.1 & 
0.785	$\pm$	0.031	$\pm$	0.068	&	0.528	$\pm$	0.017	$\pm$	0.047	&	0.311	$\pm$	0.020	$\pm$	0.032
\\
0.1 - 0.25 &
0.733	$\pm$	0.031	$\pm$	0.063	&	0.499	$\pm$	0.018	$\pm$	0.043	&	0.280	$\pm$	0.020	$\pm$	0.026
\\
0.25 - 0.4 &
0.667	$\pm$	0.032	$\pm$	0.058	&	0.433	$\pm$	0.018	$\pm$	0.037	&	0.269	$\pm$	0.025	$\pm$	0.031
\\
0.4 - 0.55 &
0.689	$\pm$	0.044	$\pm$	0.060	&	0.475	$\pm$	0.025	$\pm$	0.041	&	0.254	$\pm$	0.033	$\pm$	0.024
\\
0.55 - 0.75 &
1.218	$\pm$	0.104	$\pm$	0.106	&	0.878	$\pm$	0.059	$\pm$	0.075	&	0.508	$\pm$	0.097	$\pm$	0.044	\\
\\
0.75 - 1.5 &
-	&	-	&	-	
\\ \hline \hline
\multirow{2}{*}{$p_{T}^{2}$-bin} & &  3.2 $< \nu <$ 3.73, 0.8 $< z <$ 1.0 &  \\ \cline{2-4}
 & Carbon & Iron & Lead\\ \hline
0. - 0.1 & 
0.796	$\pm$	0.048	$\pm$	0.075	&	0.483	$\pm$	0.025	$\pm$	0.046	&	0.258	$\pm$	0.027	$\pm$	0.054
\\
0.1 - 0.25 &
0.721	$\pm$	0.038	$\pm$	0.066	&	0.472	$\pm$	0.021	$\pm$	0.044	&	0.259	$\pm$	0.028	$\pm$	0.030
\\
0.25 - 0.4 &
0.714	$\pm$	0.043	$\pm$	0.068	&	0.446	$\pm$	0.024	$\pm$	0.052	&	0.263	$\pm$	0.028	$\pm$	0.028
\\
0.4 - 0.55 &
0.777	$\pm$	0.055	$\pm$	0.074	&	0.528	$\pm$	0.030	$\pm$	0.048	&	0.210	$\pm$	0.029	$\pm$	0.026
\\
0.55 - 0.75 &
0.702	$\pm$	0.055	$\pm$	0.072	&	0.436	$\pm$	0.029	$\pm$	0.042	&	0.264	$\pm$	0.051	$\pm$	0.036
\\
0.75 - 1.5 &
-	&	-	&	-	\\ \hline \hline
\multirow{2}{*}{$p_{T}^{2}$-bin} & &  3.73 $< \nu <$ 4.2, 0.8 $< z <$ 1.0 &  \\ \cline{2-4}
 & Carbon & Iron & Lead\\ \hline
0. - 0.1 & 
0.614	$\pm$	0.119	$\pm$	0.071	&	0.337	$\pm$	0.060	$\pm$	0.093	&	0.140	$\pm$	0.058	$\pm$	0.031
\\
0.1 - 0.25 &
0.760	$\pm$	0.069	$\pm$	0.050	&	0.439	$\pm$	0.033	$\pm$	0.040	&	0.204	$\pm$	0.042	$\pm$	0.036
\\
0.25 - 0.4 &
0.701	$\pm$	0.066	$\pm$	0.057	&	0.439	$\pm$	0.033	$\pm$	0.039	&	0.149	$\pm$	0.029	$\pm$	0.037
\\
0.4 - 0.55 &
0.707	$\pm$	0.065	$\pm$	0.048	&	0.433	$\pm$	0.035	$\pm$	0.039	&	0.164	$\pm$	0.043	$\pm$	0.041
\\
0.55 - 0.75 &
0.625	$\pm$	0.073	$\pm$	0.057	&	0.433	$\pm$	0.038	$\pm$	0.038	&	0.248	$\pm$	0.067	$\pm$	0.036
\\
0.75 - 1.5 &
-	&	-	&	-	
\\ \hline \hline
\end{tabular}
\label{tab:RvzpT2}
\end{table}

\begin{table}[h!]
\caption{\label{tab:Rqvz} Numerical results for $\pi^{0}$ multiplicity ratios for C, Fe and Pb in ($Q^{2}$,~$\nu$,~$z$) bins as depicted in Fig.~{\color{blue}2}.
}
\centering
\hskip-1.0cm \begin{tabular}{c | c  c  c}
\hline \hline
\multirow{2}{*}{$z$-bin} & &  1.0 $ < $ Q$^{2} < $ 1.33, 2.2 $< \nu < $ 3.2 &  \\ \cline{2-4}
 & Carbon & Iron & Lead\\ \hline
0.3 - 0.4 & 
0.877 $\pm$ 0.020 $\pm$ 0.080 & 0.630 $\pm$ 0.012 $\pm$ 0.026 & 0.459 $\pm$ 0.013 $\pm$ 0.024 
\\
0.4 - 0.5 & 
0.838 $\pm$ 0.019 $\pm$ 0.036 & 0.618 $\pm$ 0.012 $\pm$ 0.026 & 0.471 $\pm$ 0.014 $\pm$ 0.024
\\
0.5 - 0.6 & 
0.864 $\pm$ 0.023 $\pm$ 0.038 & 0.610 $\pm$ 0.013 $\pm$ 0.026 & 0.435 $\pm$ 0.015 $\pm$ 0.030
\\
0.6 - 0.7 & 
0.881 $\pm$ 0.028 $\pm$ 0.039 & 0.642 $\pm$ 0.017 $\pm$ 0.027 & 0.448 $\pm$ 0.019 $\pm$ 0.025
\\
0.7 - 0.8 & 
0.825 $\pm$ 0.032 $\pm$ 0.041 & 0.581 $\pm$ 0.018 $\pm$ 0.027 & 0.389 $\pm$ 0.023 $\pm$ 0.022
\\
0.8 - 1.0 & 
0.715 $\pm$ 0.029 $\pm$ 0.037 & 0.514 $\pm$ 0.017 $\pm$ 0.026 & 0.326 $\pm$ 0.021 $\pm$ 0.020
\\ \hline \hline
\multirow{2}{*}{$z$-bin} & &  1.0 $ < $ Q$^{2} < $ 1.33, 3.2 $< \nu < $ 3.73 &  \\ \cline{2-4}
 & Carbon & Iron & Lead\\ \hline 
0.3 - 0.4 & 
0.867 $\pm$ 0.018 $\pm$ 0.041 & 0.664 $\pm$ 0.011 $\pm$ 0.030 & 0.501 $\pm$ 0.013 $\pm$ 0.028
\\
0.4 - 0.5 & 
0.873 $\pm$ 0.021 $\pm$ 0.041 & 0.643 $\pm$ 0.012 $\pm$ 0.030 & 0.491 $\pm$ 0.015 $\pm$ 0.029
\\
0.5 - 0.6 & 
0.838 $\pm$ 0.026 $\pm$ 0.039 & 0.642 $\pm$ 0.016 $\pm$ 0.031 & 0.440 $\pm$ 0.017 $\pm$ 0.037
\\
0.6 - 0.7 & 
0.900 $\pm$ 0.033 $\pm$ 0.042 & 0.659 $\pm$ 0.020 $\pm$ 0.035 & 0.460 $\pm$ 0.024 $\pm$ 0.040
\\
0.7 - 0.8 & 
0.891 $\pm$ 0.048 $\pm$ 0.048 & 0.658 $\pm$ 0.027 $\pm$ 0.031 & 0.488 $\pm$ 0.033 $\pm$ 0.030
\\
0.8 - 1.0 & 
0.812 $\pm$ 0.047 $\pm$ 0.046 & 0.523 $\pm$ 0.025 $\pm$ 0.027 & 0.270 $\pm$ 0.025 $\pm$ 0.024

\\ \hline \hline
\multirow{2}{*}{$z$-bin} & &  1.0 $ < $ Q$^{2} < $ 1.33, 3.73 $< \nu < $ 4.2 &  \\ \cline{2-4}
 & Carbon & Iron & Lead\\ \hline 
0.3 - 0.4 & 
0.826 $\pm$ 0.017 $\pm$ 0.037 & 0.624 $\pm$ 0.010 $\pm$ 0.028 & 0.484 $\pm$ 0.012 $\pm$ 0.027
\\
0.4 - 0.5 & 
0.830 $\pm$ 0.023 $\pm$ 0.038 & 0.628 $\pm$ 0.013 $\pm$ 0.028 & 0.489 $\pm$ 0.016 $\pm$ 0.033
\\
0.5 - 0.6 & 
0.829 $\pm$ 0.030 $\pm$ 0.039 & 0.632 $\pm$ 0.018 $\pm$ 0.028 & 0.496 $\pm$ 0.020 $\pm$ 0.032
\\
0.6 - 0.7 & 
0.947 $\pm$ 0.041 $\pm$ 0.042 & 0.632 $\pm$ 0.023 $\pm$ 0.028 & 0.412 $\pm$ 0.026 $\pm$ 0.032
\\
0.7 - 0.8 & 
0.856 $\pm$ 0.055 $\pm$ 0.039 & 0.559 $\pm$ 0.030 $\pm$ 0.030 & 0.349 $\pm$ 0.035 $\pm$ 0.023
\\
0.8 - 1.0 & 
0.751 $\pm$ 0.060 $\pm$ 0.036 & 0.434 $\pm$ 0.028 $\pm$ 0.027 & 0.228 $\pm$ 0.038 $\pm$ 0.020
\\ \hline \hline
\multirow{2}{*}{$z$-bin} & &  1.33 $ < $ Q$^{2} < $ 1.76, 2.2 $< \nu < $ 3.2 &  \\ \cline{2-4}
 & Carbon & Iron & Lead\\ \hline
0.3 - 0.4 & 
0.843 $\pm$ 0.017 $\pm$ 0.037 & 0.627 $\pm$ 0.011 $\pm$ 0.025 & 0.408 $\pm$ 0.010 $\pm$ 0.022
\\
0.4 - 0.5 & 
0.869 $\pm$ 0.017 $\pm$ 0.038 & 0.701 $\pm$ 0.011 $\pm$ 0.044 & 0.463 $\pm$ 0.012 $\pm$ 0.024
\\
0.5 - 0.6 & 
0.901 $\pm$ 0.021 $\pm$ 0.038 & 0.642 $\pm$ 0.012 $\pm$ 0.026 & 0.453 $\pm$ 0.013 $\pm$ 0.023
\\
0.6 - 0.7 & 
0.856 $\pm$ 0.023 $\pm$ 0.037 & 0.642 $\pm$ 0.014 $\pm$ 0.026 & 0.448 $\pm$ 0.016 $\pm$ 0.026
\\
0.7 - 0.8 & 
0.836 $\pm$ 0.027 $\pm$ 0.038 & 0.596 $\pm$ 0.015 $\pm$ 0.028 & 0.375 $\pm$ 0.017 $\pm$ 0.023 
\\
0.8 - 1.0 & 
0.749 $\pm$ 0.024 $\pm$ 0.035 & 0.503 $\pm$ 0.013 $\pm$ 0.026 & 0.316 $\pm$ 0.015 $\pm$ 0.021
\\ \hline \hline
\multirow{2}{*}{$z$-bin} & &  1.33 $ < $ Q$^{2} < $ 1.76, 3.2 $< \nu < $ 3.73 &  \\ \cline{2-4}
 & Carbon & Iron & Lead\\ \hline 
0.3 - 0.4 & 
0.879 $\pm$ 0.017 $\pm$ 0.043 & 0.673 $\pm$ 0.011 $\pm$ 0.030 & 0.550 $\pm$ 0.013 $\pm$ 0.029
\\
0.4 - 0.5 & 
0.850 $\pm$ 0.019 $\pm$ 0.041 & 0.654 $\pm$ 0.012 $\pm$ 0.030 & 0.499 $\pm$ 0.014 $\pm$ 0.027 
\\
0.5 - 0.6 & 
0.900 $\pm$ 0.025 $\pm$ 0.042 & 0.634 $\pm$ 0.014 $\pm$ 0.029 & 0.518 $\pm$ 0.018 $\pm$ 0.035
\\
0.6 - 0.7 & 
0.891 $\pm$ 0.031 $\pm$ 0.042 & 0.617 $\pm$ 0.017 $\pm$ 0.029 & 0.503 $\pm$ 0.022 $\pm$ 0.035
\\
0.7 - 0.8 & 
0.904 $\pm$ 0.040 $\pm$ 0.046 & 0.649 $\pm$ 0.023 $\pm$ 0.032 & 0.409 $\pm$ 0.027 $\pm$ 0.024
\\
0.8 - 1.0 & 
0.815 $\pm$ 0.038 $\pm$ 0.042 & 0.537 $\pm$ 0.021 $\pm$ 0.029 & 0.337 $\pm$ 0.026 $\pm$ 0.021
\\ \hline \hline
\multirow{2}{*}{$z$-bin} & &  1.33 $ < $ Q$^{2} < $ 1.76, 3.73 $< \nu < $ 4.2 &  \\ \cline{2-4}
 & Carbon & Iron & Lead\\ \hline 
0.3 - 0.4 & 
0.829 $\pm$ 0.017 $\pm$ 0.069 & 0.638 $\pm$ 0.011 $\pm$ 0.030 & 0.474 $\pm$ 0.012 $\pm$ 0.029
\\
0.4 - 0.5 & 
0.911 $\pm$ 0.024 $\pm$ 0.044 & 0.646 $\pm$ 0.014 $\pm$ 0.030 & 0.511 $\pm$ 0.016 $\pm$ 0.031
\\
0.5 - 0.6 & 
0.915 $\pm$ 0.032 $\pm$ 0.044 & 0.625 $\pm$ 0.018 $\pm$ 0.032 & 0.467 $\pm$ 0.020 $\pm$ 0.031
\\
0.6 - 0.7 & 
0.935 $\pm$ 0.039 $\pm$ 0.045 & 0.679 $\pm$ 0.023 $\pm$ 0.033 & 0.433 $\pm$ 0.026 $\pm$ 0.037
\\
0.7 - 0.8 & 
0.843 $\pm$ 0.049 $\pm$ 0.055 & 0.598 $\pm$ 0.029 $\pm$ 0.035 & 0.383 $\pm$ 0.038 $\pm$ 0.031
\\
0.8 - 1.0 & 
0.672 $\pm$ 0.050 $\pm$ 0.052 & 0.441 $\pm$ 0.025 $\pm$ 0.032 & 0.275 $\pm$ 0.036 $\pm$ 0.029
\\ \hline \hline
\multirow{2}{*}{$z$-bin} & &  1.76 $ < $ Q$^{2} < $ 4.1, 2.2 $< \nu < $ 3.2 &  \\ \cline{2-4}
 & Carbon & Iron & Lead\\ \hline
0.3 - 0.4 & 
0.915 $\pm$ 0.023 $\pm$ 0.046 & 0.669 $\pm$ 0.014 $\pm$ 0.035 & 0.534 $\pm$ 0.017 $\pm$ 0.036
\\
0.4 - 0.5 & 
0.869 $\pm$ 0.022 $\pm$ 0.061 & 0.660 $\pm$ 0.014 $\pm$ 0.036 & 0.486 $\pm$ 0.016 $\pm$ 0.035
\\
0.5 - 0.6 & 
0.902 $\pm$ 0.027 $\pm$ 0.045 & 0.674 $\pm$ 0.016 $\pm$ 0.035 & 0.469 $\pm$ 0.018 $\pm$ 0.035
\\
0.6 - 0.7 & 
0.901 $\pm$ 0.032 $\pm$ 0.046 & 0.693 $\pm$ 0.020 $\pm$ 0.037 & 0.504 $\pm$ 0.023 $\pm$ 0.040
\\
0.7 - 0.8 & 
0.836 $\pm$ 0.033 $\pm$ 0.045 & 0.629 $\pm$ 0.020 $\pm$ 0.035 & 0.325 $\pm$ 0.022 $\pm$ 0.037
\\
0.8 - 1.0 & 
0.772 $\pm$ 0.031 $\pm$ 0.043 & 0.543 $\pm$ 0.018 $\pm$ 0.033 & 0.328 $\pm$ 0.021 $\pm$ 0.038
\\ \hline \hline
(Continued)
\end{tabular}
\end{table}

\begin{table}[h]
\centering
\hskip-1.0cm \begin{tabular}{c | c  c  c}
\hline \hline
\multirow{2}{*}{$z$-bin} & &  1.76 $ < $ Q$^{2} < $ 4.1, 3.2 $< \nu < $ 3.73 &  \\ \cline{2-4}
 & Carbon & Iron & Lead\\ \hline 
0.3 - 0.4 & 
0.917 $\pm$ 0.016 $\pm$ 0.042 & 0.689 $\pm$ 0.010 $\pm$ 0.032 & 0.517 $\pm$ 0.011 $\pm$ 0.031
\\
0.4 - 0.5 & 
0.876 $\pm$ 0.017 $\pm$ 0.041 & 0.651 $\pm$ 0.011 $\pm$ 0.031 & 0.496 $\pm$ 0.012 $\pm$ 0.030
\\
0.5 - 0.6 & 
0.894 $\pm$ 0.022 $\pm$ 0.042 & 0.683 $\pm$ 0.013 $\pm$ 0.031 & 0.500 $\pm$ 0.015 $\pm$ 0.032
\\
0.6 - 0.7 & 
0.892 $\pm$ 0.027 $\pm$ 0.044 & 0.640 $\pm$ 0.016 $\pm$ 0.031 & 0.468 $\pm$ 0.019 $\pm$ 0.030
\\
0.7 - 0.8 & 
0.796 $\pm$ 0.028 $\pm$ 0.038 & 0.615 $\pm$ 0.017 $\pm$ 0.031 & 0.443 $\pm$ 0.022 $\pm$ 0.027
\\
0.8 - 1.0 & 
0.802 $\pm$ 0.028 $\pm$ 0.038 & 0.548 $\pm$ 0.016 $\pm$ 0.030 & 0.299 $\pm$ 0.018 $\pm$ 0.024
\\ \hline \hline

\multirow{2}{*}{$z$-bin} & &  1.76 $ < $ Q$^{2} < $ 4.1, 3.73 $< \nu < $ 4.2 &  \\ \cline{2-4}
 & Carbon & Iron & Lead\\ \hline 
0.3 - 0.4 & 
0.873 $\pm$ 0.016 $\pm$ 0.040 & 0.651 $\pm$ 0.010 $\pm$ 0.029 & 0.513 $\pm$ 0.011 $\pm$ 0.029
\\
0.4 - 0.5 & 
0.894 $\pm$ 0.021 $\pm$ 0.041 & 0.661 $\pm$ 0.012 $\pm$ 0.030 & 0.512 $\pm$ 0.015 $\pm$ 0.030
\\
0.5 - 0.6 & 
0.905 $\pm$ 0.028 $\pm$ 0.045 & 0.658 $\pm$ 0.017 $\pm$ 0.030 & 0.495 $\pm$ 0.020 $\pm$ 0.041
\\
0.6 - 0.7 & 
0.905 $\pm$ 0.033 $\pm$ 0.042 & 0.656 $\pm$ 0.020 $\pm$ 0.029 & 0.461 $\pm$ 0.022 $\pm$ 0.027
\\
0.7 - 0.8 & 
0.869 $\pm$ 0.040 $\pm$ 0.042 & 0.632 $\pm$ 0.024 $\pm$ 0.029 & 0.362 $\pm$ 0.028 $\pm$ 0.030
\\
0.8 - 1.0 & 
0.733 $\pm$ 0.038 $\pm$ 0.037 & 0.566 $\pm$ 0.023 $\pm$ 0.027 & 0.273 $\pm$ 0.027 $\pm$ 0.029
\\ \hline \hline

\end{tabular}
\end{table}

\end{document}